\documentclass[amsmath,amssymb, aps, prl, twocolumn, notitlepage]{revtex4-2}
\usepackage{graphicx}
\usepackage{dcolumn}
\usepackage{bm}
\usepackage[usenames]{color}
\usepackage{slashed}
\usepackage[utf8]{inputenc}
\usepackage{amsfonts}
\usepackage{amsmath}
\usepackage{amssymb}
\usepackage{hyperref}
\usepackage{latexsym}
\usepackage{color}
\usepackage{cancel}
\usepackage{braket}

\begin{document}

\title{Complex semiclassical theory for non-Hermitian quantum systems}

\author{Guang Yang\textsuperscript{\tiny $\#$}}
\affiliation{International Center for Quantum Materials, School of Physics, Peking University, Beijing, 100871, China}
\author{Yong-Kang Li\textsuperscript{\tiny $\#$}}
\affiliation{International Center for Quantum Materials, School of Physics, Peking University, Beijing, 100871, China}
\author{Yongxu Fu}
\affiliation{International Center for Quantum Materials, School of Physics, Peking University, Beijing, 100871, China}
\author{Zhenduo Wang}
\affiliation{International Center for Quantum Materials, School of Physics, Peking University, Beijing, 100871, China}
\author{Yi Zhang}
\email{frankzhangyi@gmail.com}
\thanks{\textsuperscript{\tiny $\#$}G. Yang and Y.-K. Li contributed equally.}
\affiliation{International Center for Quantum Materials, School of Physics, Peking University, Beijing, 100871, China}

\begin{abstract}
Non-Hermitian quantum systems exhibit fascinating characteristics such as non-Hermitian topological phenomena and skin effect, yet their studies are limited by the intrinsic difficulties associated with their eigenvalue problems, especially in larger systems and higher dimensions. In Hermitian systems, the semiclassical theory has played an active role in analyzing spectrum, eigenstate, phase, transport properties, etc. Here, we establish a complex semiclassical theory applicable to non-Hermitian quantum systems by an analytical continuation of the physical variables such as momentum, position, time, and energy in the equations of motion and quantization condition to the complex domain. Further, we propose a closed-orbit scheme and physical meaning under such complex variables. We demonstrate that such a framework straightforwardly yields complex energy spectra and quantum states, topological phases and transitions, and even the skin effect in non-Hermitian quantum systems, presenting an unprecedented perspective toward nontrivial non-Hermitian physics, even with larger systems and higher dimensions.
\end{abstract}

\maketitle

\emph{Introduction} \textemdash The non-Hermitian skin effect (NHSE), where most eigenstates localize at open boundaries, reveals that the physics of non-Hermitian quantum systems may differ significantly from their Hermitian counterparts \cite{Wang2018, Emil2018, Murakami2019, Slager2020, Fang2020, Sato2020, Fanghu2020, bergholtzrev2021, ashida2020}. A series of nontrivial non-Hermitian topological phases have also been established as generalizations of topological phenomena \cite{SatoUedaPRX2019, UedaPRX2018, WangPRL2018NHChernband, WangPRL2019, LieuPRB2018, LongshiPRL2019, UedaNoriPRL2019, SongPRB2019, xiao2020non, FuPRL2018,ghatak2019new, Sato2023, NoriPRL2017, Yongxu2021NHSE, MaFangNRP2022, Yongxu2022deg, Yongxu2023bulk}. Further studies on the non-Hermitian emergence, such as dissipation in quantum optics \cite{Ozdemir2019, Ganainy2018, Chen2017, Feng2017, AluScience2019, RMP2019, HenningPRL2015, LiNC2020CNHSE, FangPRL2022NHSEin3d, MahitoPRB2011, LeePRL2016}, open systems in cold atom systems \cite{Zoller2008, XueNP2020NHSEinAMO, GongPRL2020NHSEinAMO, UedaPRL2018AMO, DuanPRL2017NHAMO}, and finite quasiparticle lifetime in condensed matters \cite{kozii2017non, FuPRL2018QO, FuPRB2019, NorioPRB2018}, etc. have brought researches on non-Hermitian quantum systems to the recent frontier.

The semiclassical theory \cite{Onsager1952, Lifshitz1956, QianNiu2010RMP} has offered straightforward analysis and understanding of physical properties in Hermitian quantum materials and models, such as spectra, transport, and topological phenomena. For example, the semiclassical framework has foreshadowed various experimental and theoretical conclusions and discoveries for the quantum Hall effects in two and three dimensions \cite{IQHE1980, TKNN1982, Laughlin1981IQH, Zhang2017, Wang2017, Zhang2018, LHL20203DQHE}, as well as the quantum anomalous \cite{QiARCMP2016QAHE, FangAIP2015QAHE, XueScience2013QAHE} and quantum spin Hall effects \cite{ZhangPRL2006QSHE, KaneMelePRL2015QSHEinG, ZhangScience2006QSHE, KaneMelePRL2005QSHE}, etc. However, despite semiclassical studies on the Berry curvature \cite{IlanPRB2020, Fan_2020}, with only a constant electric field and no magnetic field \cite{wang2022anomalous}, or only for real energy spectra \cite{XingPRB2022}, a more comprehensive semiclassical theory for non-Hermitian quantum systems is still lacking. Given the intrinsic difficulties and numerical instabilities of eigenvalue problems in non-Hermitian quantum systems \cite{Fanghu2020, Zhong2021amplification, FuPRL2018}, especially on relatively larger systems and higher dimensions, such a semiclassical theory becomes even more compelling.

In this letter, we develop a semiclassical theory, which we dub the complex semiclassical theory, for more generic non-Hermitian quantum systems, including scenarios with complex energy spectra and variable electric and magnetic fields. The equations of motion (EOMs) and quantization conditions generalize those of the Hermitian quantum systems; yet, we require analytical continuations of its physical variables, such as momentum and position, to the complex domains, where we also establish a straightforward strategy to search for the vital closed orbits. Further, we explain the physical meaning of wave packets' complex-valued variables originating from the biorthogonal basis we work with. When demonstrated on and applied to various non-Hermitian quantum systems, even in scenarios far exceeding previous system-size and dimension thresholds, the complex semiclassical theory provides us with accurate complex energy spectra, quantum eigenstates, non-Hermitian topological phase diagrams, emergence of the non-Hermitian skin effect, and more.

\emph{Complex semiclassical theory} \textemdash Conventionally, the semiclassical theory begins with the EOMs of a wave packet's center of mass (COM) in a quantum system: \cite{QianNiu2010RMP}:
\begin{eqnarray}
\dot {\bf r} &=& \partial \epsilon / \partial {\bf p} - \dot {\bf p}\times {\bf \Omega}, \nonumber\\
\dot {\bf p} &=& {\bf E} + \dot {\bf r}\times {\bf B}, \label{eq:EOM}
\end{eqnarray}
where ${\bf E} = -\partial \epsilon / \partial {\bf r}$ and ${\bf B}$ are the electric field and magnetic field, respectively. For a valid quantized energy level, we require the trajectory to form a closed orbit satisfying the Bohr-Sommerfeld quantization condition:
\begin{equation}
\oint {\bf p} \cdot d{\bf r} = (n+\gamma)h, n \in \mathbb{Z}, \label{eq:QC}
\end{equation}
where $\gamma=1/2$ and in the absence of the Berry curvature ${\bf \Omega}=0$ \cite{QianNiu2010RMP, IlanPRB2020, Fan_2020}. We study scenarios with the Berry curvature in the supplemental materials \cite{SuppCST}. As the computations of Eqs. \ref{eq:EOM} and \ref{eq:QC} are virtually  classical, they offer very efficient evaluations of the target quantum systems.

Conventionally, the variables in the EOMs such as $\bf r$, $\bf p$, and $\epsilon$ take real values, as they are expectation values of Hermitian observables and correspond to classical quantities. Interestingly, to generalize the semiclassical theory to non-Hermitian quantum systems, we analytically continue these physical variables to the complex domains \footnote{See a WKB example with ${\bf p} \in \mathbb{C}$ and ${\bf r} \in \mathbb{R}$ in Ref. \cite{SuppCST}. }, hence the name complex semiclassical theory. Following Eq. \ref{eq:EOM}, we may obtain trajectories using finite-time steps $dt$ with sufficiently small $|dt|$ to suppress finite-difference errors \cite{Runge-Kutta1, Runge-Kutta2, Runge-Kutta3, Bulirsch--Stoer1, Bulirsch--Stoer2}. It is straightforward to see that $\dot{\epsilon}=0$; thus, though complex, $\epsilon$ is constant \footnote{Correspondingly, the biorthogonal quantum expectation value $\langle\psi_L(t)|\hat{H}|\psi_R(t)\rangle=\langle\psi_L(0)|e^{i\hat{H}t}\hat{H}e^{-i\hat{H}t}|\psi_R(0)\rangle=\langle\psi_L(0)|\hat{H}|\psi_R(0)\rangle$ remains constant irrespective of $t$. }. Once we locate a closed orbit, we evaluate its geometric phase through a complex integral $\oint {\bf p} \cdot d {\bf r}$; if the quantization condition Eq. \ref{eq:QC} holds, the underlying $\epsilon$ and complex semiclassical orbit contribute to the spectrum and eigenstates and offer insights on the physical property of the non-Hermitian quantum system, as far as the semiclassical validity holds, e.g., guaranteed by integrability. We will establish these main conclusions later in the letter.

In practice, however, unlike the straightforward one-dimensional search in $t\in \mathbb{R}$ for closed orbit in the Hermitian cases, the search for closed orbit in $t\in \mathbb{C}$ in the complex semiclassical theory for non-Hermitian quantum systems may turn out aimless and challenging \cite{Cazenave1982, Runge-Kutta1, Runge-Kutta2, Runge-Kutta3, Bulirsch--Stoer1, Bulirsch--Stoer2, lyapunovstability1, lyapunovstability2}, especially since the complex period $T$ is initially unknown and may differ between $\epsilon$. To address such problems, we propose a two-step strategy: (1) we open up a trajectory that somewhat approaches the initial coordinates, then (2) choose the complex phase of subsequent $dt$ according to $\dot {\bf r}$ (or $\dot {\bf p}$) in the subsequent steps, so that $d{\bf r}$ ($d{\bf p}$) reduce $\Delta{\bf r}$ ($\Delta{\bf p}$) - the differences between initial and current coordinates - until they vanish, and we obtain a closed orbit. As complex integrals around closed loops do not depend on paths (but on windings around certain ``fixed points"), such make-shift closed orbits provide equal evaluations of vital physical quantities, such as the period $T$ and the geometric phase $\oint {\bf p} \cdot d{\bf r}$. Once $T$ is available, we may resort to smoother, more aesthetic orbits along $t \parallel T$. This approach does not require fine-tuning or perturbative treatment of the non-Hermitian terms.

\emph{Toy model examples} \textemdash As a first example, let's first consider the following quadratic model in 1D \footnote{Generalizations to cases with finite linear and constant terms are straightforward.}:
\begin{equation}
    \hat H = \alpha \hat x^2 + \beta \hat p^2 + \eta (\hat x\hat p + \hat p\hat x)/2, \label{eq:ho_Ham}
\end{equation}
where $\alpha, \beta, \eta \in \mathbb{C}$, and $\hat H$ is non-Hermitian in general. The model is exactly solvable with the ladder operators $[\hat a, \hat{b}^\dagger]=1$:
\begin{equation}
\hat a = \frac{2\alpha\hat x+(\eta+i\omega)\hat p}{2\sqrt{\alpha\omega}}, \hat b^\dagger = \frac{2\alpha\hat x+(\eta-i\omega)\hat p}{2\sqrt{\alpha\omega}}.
\end{equation}
We set $\hbar=1$ here and afterward. The energy spectrum $\epsilon_n = (n+1/2)\omega$, $n\in \mathbb{Z}$, $\omega = \sqrt{4\alpha\beta-\eta^2}$ offers a solid benchmark.

\begin{figure}
\includegraphics[width=0.49\linewidth]{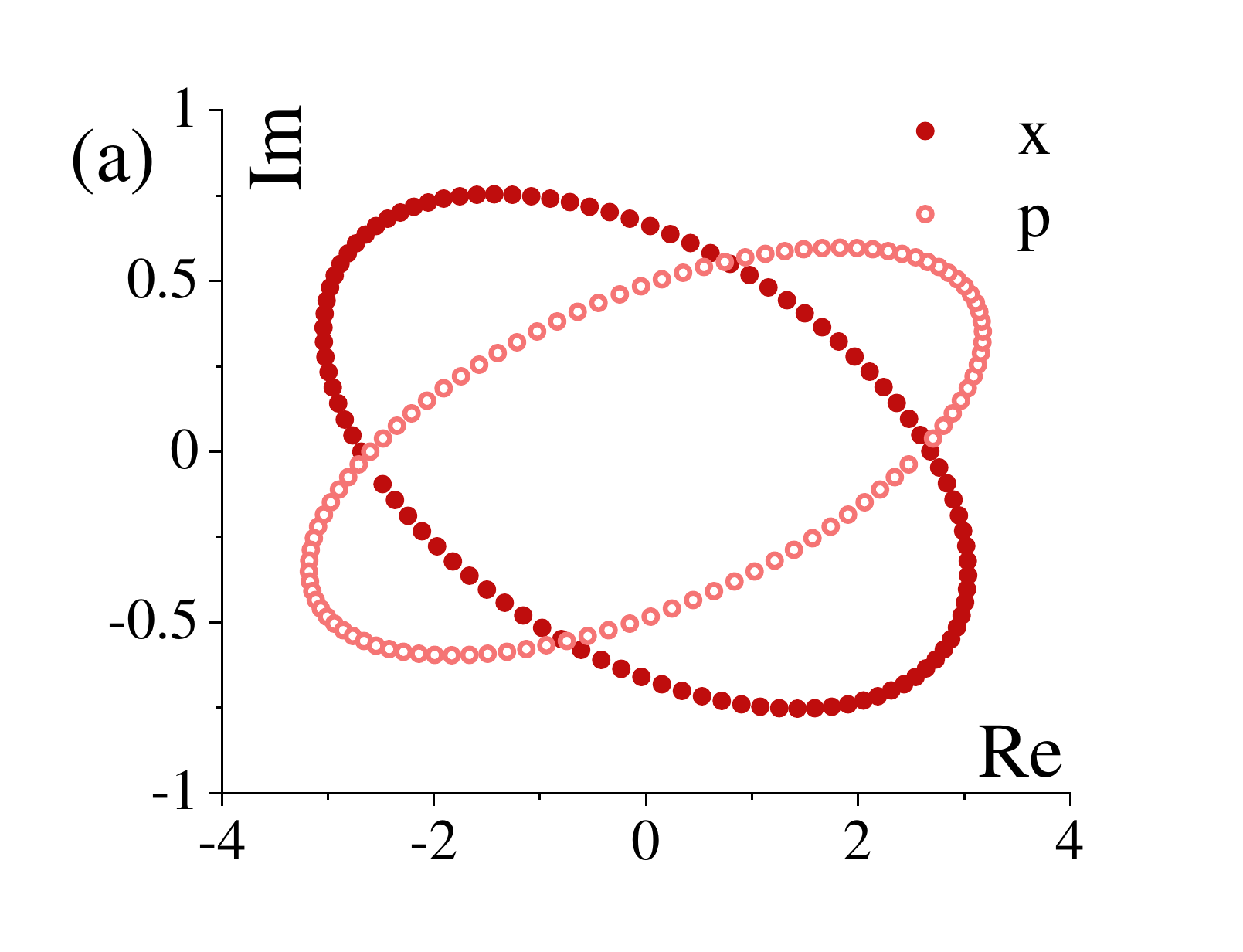}
\includegraphics[width=0.49\linewidth]{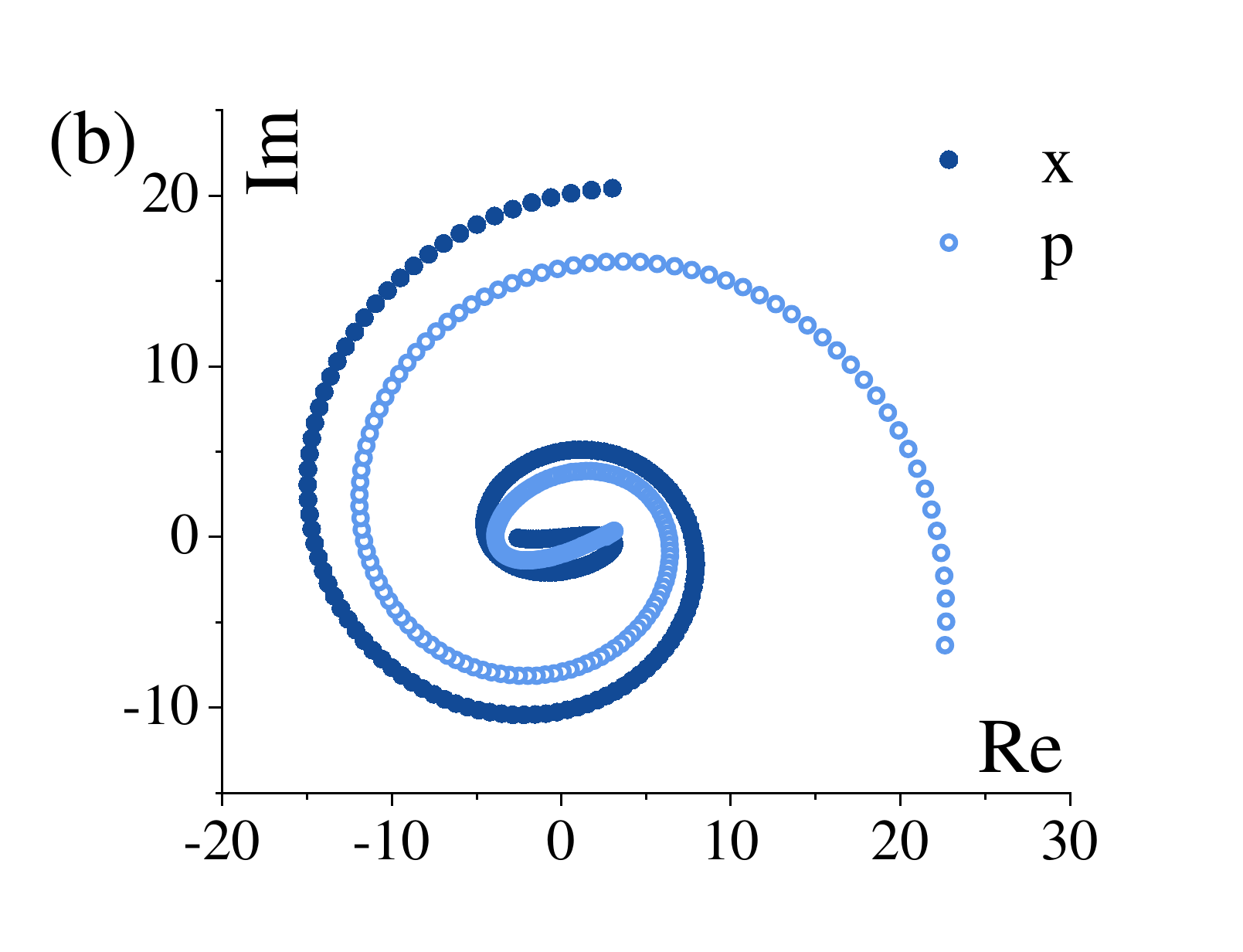}
\includegraphics[width=0.49\linewidth]{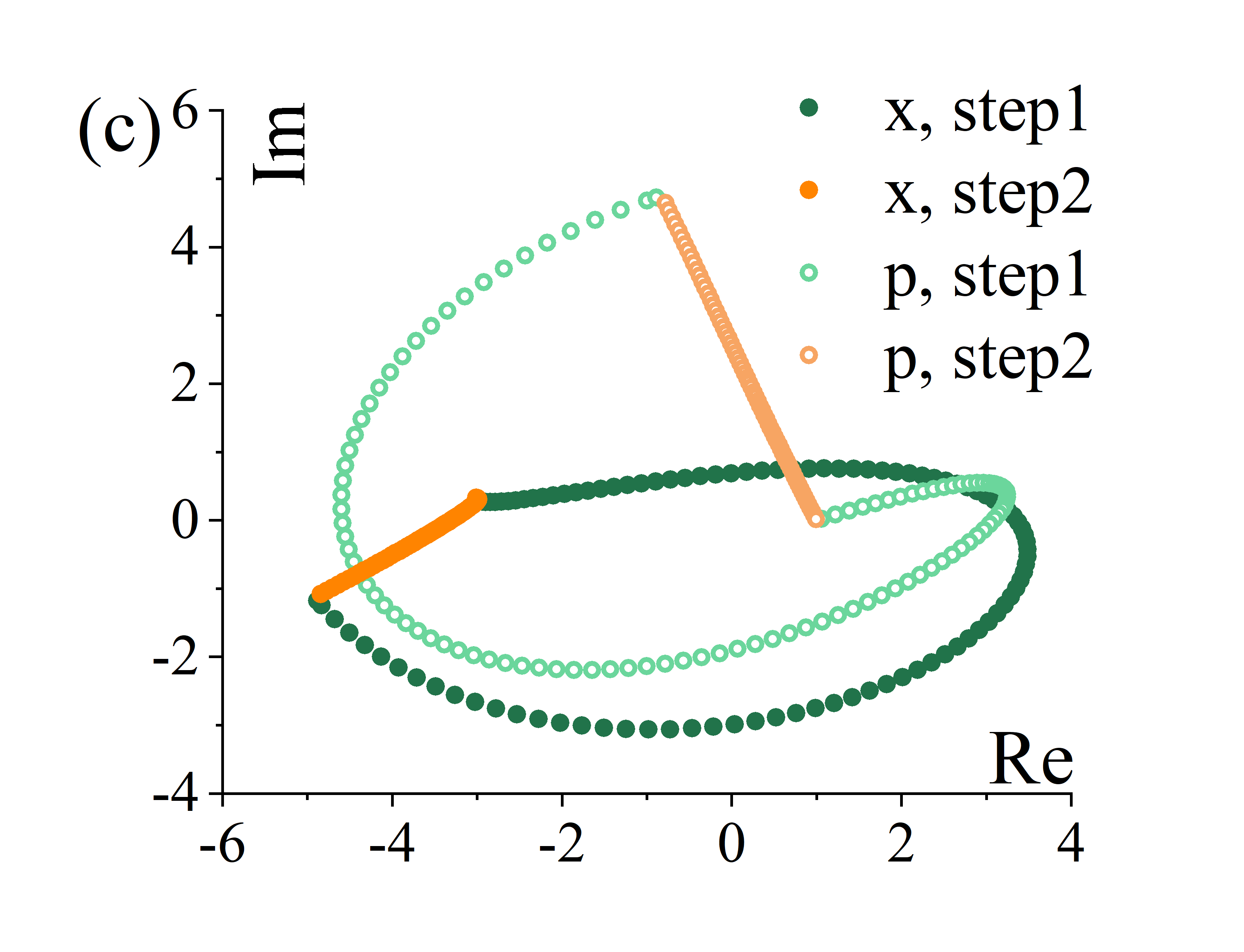}
\includegraphics[width=0.49\linewidth]{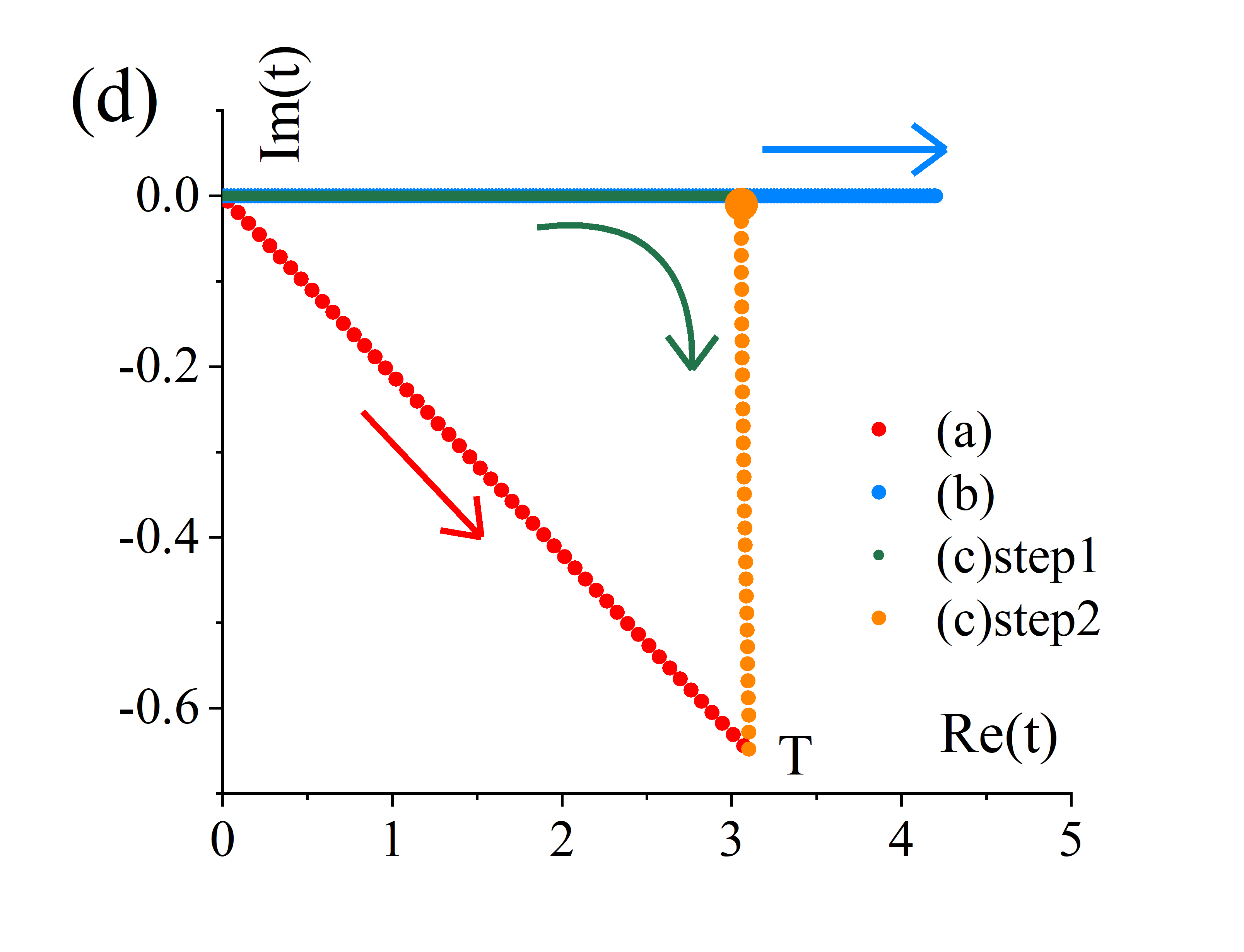}
\caption{Following the semiclassical EOMs of the model in Eq. \ref{eq:ho_Ham}, the trajectories form (a) a closed loop for $\omega t\in \mathbb{R}$, or (b) an open spiral for $t \in \mathbb{R}$. (c) Without knowledge of $T$, we can still obtain a closed loop from an open trajectory (step 1) followed by finite-time steps (step 2) with $dt$ in Eq. \ref{eq:dt}. (d) The trajectories of different closed orbits in the complex $t$ plane reach the same complex period $T=2\pi/\omega$. $\alpha = 1+0.5i$, $\beta=1$, and $\eta = 0.7+0.3i$. }
\label{fig:HOorbit}
\end{figure}

Eq. \ref{eq:ho_Ham} is beyond previous semiclassical frameworks. From the complex semiclassical theory, we start with the complex energy $\epsilon = \alpha x^2 + \beta p^2 + \eta xp$, and the corresponding EOMs following Eq. \ref{eq:EOM} yields the following trajectories \footnote{An alternative, heuristic derivation from a wave-packet perspective is in Ref. \cite{SuppCST}.}:
\begin{eqnarray}
x(t) &=& z_1 e^{i\omega t} + z_2 e^{-i\omega t},  \nonumber\\
p(t) &=& \frac{(i\omega-\eta)z_1}{2\beta} e^{i\omega t} - \frac{(i\omega+\eta)z_2}{2\beta} e^{-i\omega t},
\label{eq:qdr_EOM}
\end{eqnarray}
where $z_1, z_2\in \mathbb{C}$ are determined by the initial conditions. Eq. \ref{eq:qdr_EOM} forms closed orbits (Fig. \ref{fig:HOorbit}a) with a generally complex-valued period $T=2\pi/\omega$. However, this is a privilege for solvable trajectories, which otherwise end up as open spirals (Fig. \ref{fig:HOorbit}b) for generic $t$, e.g., trajectories from finite-time steps. Fortunately, we can follow our strategy to guarantee closed orbits (Fig. \ref{fig:HOorbit}c). First, we choose a trajectory that, to some degree, circles near the initial point; then, we set $dt$ for the subsequent steps as:
\begin{equation}
dt = \frac{[x(0)-x(t)]/\dot{x}}{|[x(0)-x(t)]/\dot{x}|}|dt|, \label{eq:dt}
\end{equation}
which ensures that $dx = \dot{x}dt = C[x(0)-x(t)]$, $C>0$, and reduces $|x(0)-x(t)|$, until $x(t)$ returns to its initial value $x(0)$ \footnote{The constraint of constant $\epsilon$ guarantees $p$ return to its initial value as well. Similarly, we can pick $p$ as our criteria for $dt$ with the same effect. }.

Finally, we evaluate physical quantities over the closed orbits: $\epsilon= z_1 z_2 \omega^2 /\beta $ and $\oint p \cdot dx = 2\pi z_1 z_2 \omega / \beta$. Thus, the quantization condition dictates $z_1 z_2 = (n+1/2)\beta/\omega$, and in turn, the energy spectrum to quantize as $\epsilon_n = (n+1/2)\omega$, consistent with the benchmark. We also examine continuous model examples with higher-order terms in the supplemental materials \cite{SuppCST}.

\emph{Derivation and interpretation} \textemdash The semiclassical theory is based upon a wave-packet representation of quantum states. Given the center-of-mass (COM) position ${\bf r}_c$ and momentum ${\bf p}_c$ of a wave packet, we define a corresponding quantum state $|{\bf r}_c, {\bf p}_c\rangle =\hat W({\bf r}_c, {\bf p}_c)|0\rangle$, where $\hat W({\bf r}_c, {\bf p}_c) = \exp[i({\bf p}_c\cdot \hat{\bf r}-{\bf r}_c \cdot \hat{\bf p})]$ is a translation operator and $|0\rangle$ is a wave packet centered at zero position and momentum. We can also obtain ${\bf r}_c$ and ${\bf p}_c$ from $|{\bf r}_c, {\bf p}_c\rangle$ through its expectation values:
\begin{eqnarray}
\langle {\bf r}_c, {\bf p}_c| \hat{\bf r}|{\bf r}_c, {\bf p}_c\rangle = {\bf r}_c, \; \langle {\bf r}_c, {\bf p}_c| \hat{\bf p}|{\bf r}_c, {\bf p}_c\rangle = {\bf p}_c, \label{eq:op_wavepacket}
\end{eqnarray}
thus establishing a mapping between a wave packet $|{\bf r}_c, {\bf p}_c\rangle$ and its physical variables ${\bf r}_c$ and ${\bf p}_c$.

For the complex semiclassical theory, we can generalize ${\bf r}_c, {\bf p}_c \in \mathbb{C}$, which however, will require the biorthogonal basis with $\langle {\bf r}_c, {\bf p}_c| = \langle 0| \hat W(-{\bf r}_c, -{\bf p}_c)$, generally different from $\langle 0| \hat W({\bf r}_c, {\bf p}_c)^\dagger$. Likewise, an observable $\hat O$'s expectation value evolves as:
\begin{equation}
\langle \hat O\rangle (t) = \langle {\bf r}_c, {\bf p}_c| U^{-1}(t) \hat O U(t)|{\bf r}_c, {\bf p}_c\rangle,
\label{eq:complex_ev}
\end{equation}
where $U(t) = \exp(-i\hat H t)$. Note on the bra side, we have employed the inverse $\langle {\bf r}_c, {\bf p}_c|U^{-1}(t)$ instead of the Hermitian conjugate $\langle {\bf r}_c, {\bf p}_c|U^{\dagger}(t)$. Such a biorthogonal basis $\langle nL|mR\rangle = \delta_{nm}$ \footnote{The expectation value and time evolution operator under the biorthogonal basis are $\langle nL|\hat{O}|nR\rangle$ and $U(t)=\sum_n e^{-iE_n t}|nR\rangle\langle nL|$ \cite{Haoshu2022}, respectively.} is advantageous in keeping the normalization constant $\langle {\bf r}_c, {\bf p}_c| U^{-1}(t) \cdot \mathbf{1}\cdot U(t)|{\bf r}_c, {\bf p}_c\rangle = 1$ and the operator evolution $\dot {\hat O} = -i[\hat O, \hat H]$ in the Heisenberg picture meaningful. However, there is a trade-off: even for a Hermitian $\hat O$, $\dot {\hat O}$ is no longer necessarily Hermitian under a non-Hermitian $\hat H$, hence the analytical continuation of the expectation values, e.g., Eq. \ref{eq:complex_ev}, and the physical variables to the complex domains. Our biorthogonal-basis convention differs from its common alternative, i.e., $\langle \Psi | U^{\dagger}(t) \hat O U(t)|\Psi\rangle$, which keeps expectation values real-valued and focuses on solely one set of (right) eigenstates, thus suitable for analyzing their properties and quantum dynamics \cite{Husimi2023}. These two conventions coincide for a Hermitian $\hat{H}$.

Next, we analyze the evolution of such complex variables, following the wave packet $|{\bf r}_c(t+\tau), {\bf p}_c(t+\tau)\rangle = e^{-i\hat{H}\tau}|{\bf r}_c(t), {\bf p}_c(t)\rangle$ after a short time step $\tau \rightarrow 0$:
\begin{eqnarray}
\dot{\bf r}_c(t) &=& \left[{\bf r}_c(t+\tau) - {\bf r}_c(t)\right]/\tau \label{eq:TR_deri} \nonumber\\
&=& \langle {\bf r}_c(t), {\bf p}_c(t)| e^{i\hat H \tau} \hat{\bf r} e^{-i\hat{H}\tau} - \hat{\bf r}|{\bf r}_c(t), {\bf p}_c(t) \rangle / \tau \\
&=& -i\langle 0| \hat W(-{\bf r}_c(t), -{\bf p}_c(t)) \left[\hat{\bf r}, \hat{H}\right] \hat W({\bf r}_c(t), {\bf p}_c(t))|0\rangle \nonumber \\
&=& \langle 0|\frac{\partial \hat H}{\partial \hat{\bf p}}(\hat{\bf r}+{\bf r}_c(t), \hat{\bf p}+{\bf p}_c(t))|0\rangle \approx \partial \epsilon({\bf p}_c, {\bf r}_c)/\partial {\bf p}_c, \nonumber
\end{eqnarray}
which yields the EOM for the complex variable ${\bf r}_c$ in Eq. \ref{eq:EOM}. Similarly, we obtain the EOM for ${\bf p}_c$ as well as the case with $\bf \Omega$ \cite{SuppCST}. For the last step, we have kept only the $m=n=0$ term in the Taylor expansion (equivalent to replacing operators $\hat{\bf r}$ and $\hat{\bf p}$ in $\partial\hat{H}/\partial{\bf p}$ with complex variables ${\bf r}_c$ and ${\bf p}_c$):
\begin{equation}
\frac{\partial \hat H}{\partial \hat{\bf p}}(\hat{\bf r}+{\bf r}_c(t), \hat{\bf p}+{\bf p}_c(t)) = \sum_{mn}\frac{\hat{\bf r}^n \hat{\bf p}^m}{n!m!}\frac{\partial \hat{H}}{\partial^n \hat{\bf r} \partial^{m+1} \hat{\bf p}}({\bf r}_c(t), {\bf p}_c(t)).
\end{equation}
The higher-order terms are negligible because (1) the first order terms, i.e., $\langle 0|\hat{\bf r}|0 \rangle$ and $\langle 0|\hat{\bf p}|0 \rangle$ vanish due to the COM definition of wave packet $|0\rangle$, and (2) the second and higher order terms are negligible in comparison with their classical counterparts, e.g., $\langle 0|\hat{\bf r}^2|0 \rangle$ versus ${\bf r}_c^2$, as the extent of the wave packet is small in comparison with the orbit - a prerequisite for semiclassical approximation that holds at large quantum numbers. Further, the approximation becomes exact for a quadratic $\hat{H}$, where higher-order terms vanish. We also show scenarios where such approximation becomes mediocre in the quantum limit and with influential higher orders in Ref. \cite{SuppCST}.

\begin{table}[]
\begin{tabular}{|l|l|l|l|}
\hline
$n$ & Quantum      & Semiclassical     & Outcome      \\\hline
0 & $|0\rangle_0$    & $|0\rangle_0$         & $|0\rangle_0$    \\\hline
1 & $(\hat a^\dagger_0+\kappa \hat a_0)|0\rangle_0$   & $\hat a^\dagger_0|0\rangle_0$    & $|1\rangle_0$            \\\hline
2 & $(\hat a^\dagger_0+\kappa \hat a_0)^2|0\rangle_0$ & $[(\hat a^\dagger_0)^2+\kappa]|0\rangle_0$ & $\sqrt{2}|2\rangle_0 +\kappa|0\rangle_0$ \\\hline
3 & $(\hat a_0^\dagger+\kappa \hat a_0)^3|0\rangle_0$ & $[(\hat a^\dagger_0)^3+3\kappa \hat{a}_0^\dagger]|0\rangle_0$     & $\sqrt{6}|3\rangle_0+3\kappa|1\rangle_0$ \\\hline
\end{tabular}
\caption{The lowest, non-orthogonal eigenstates of the non-Hermitian quantum system in Eq. \ref{eq:ho_Ham} obtained via quantum and semiclassical approaches compare consistently \cite{SuppCST}. For clarity, we have employed the orthonormal basis $|m\rangle_0$ of a Hermitian, isotropic harmonic oscillator: $a^\dagger_0 a_0|m\rangle_0 = m|m\rangle_0$, $\hat a_0 = (\hat x+i\hat p)/\sqrt{2}$. $2\alpha=\omega-i\eta$, $2\beta=\omega+i\eta$, $\kappa = \eta/i\omega$. }
\label{tab:statecompare}
\end{table}

What is the physical meaning of such a wave packet with complex variables? To demonstrate, we define $x'_c=\Re(x_c)$ and $x''_c=\Im(x_c)$ [$p'_c=\Re(p_c)$ and $p''_c=\Im(p_c$)] as the real and imaginary parts of $x_c$ ($p_c$) in 1D, then $|x_c, p_c \rangle$:
\begin{equation}
W(x_c, p_c)|0\rangle = \mathcal{A}(x_c, p_c) W(x_r, p_r)|0\rangle = \mathcal{A}(x_c, p_c) |x_r, p_r \rangle, \label{eq:coh_map2}
\end{equation}
is essentially a wave packet with real-valued COM position $x_r = x'_c -p''_c\in \mathbb{R}$ and momentum $p_r=p'_c+x''_c\in \mathbb{R}$, yet with an extra factor:
\begin{eqnarray}
\mathcal{A}(x_c, p_c)&=&\exp(\frac{x_r^2+p_r^2-x_c^2-p_c^2}{4}) \in \mathbb{C}.
\label{eq:coh_map}
\end{eqnarray}
Indeed, given the non-unitary evolution in non-Hermitian quantum systems, it is natural to encompass variable amplitudes (and phases).

We can also convert the real-valued COM $x_r, p_r\in \mathbb{R}$ and overall factor $\mathcal{A}\in \mathbb{C}$ into $x_c, p_c\in \mathbb{C}$, as both parameterizations employ four real numbers. Similarly, upon changes of $\mathcal{A}\rightarrow \mathcal{A}\cdot \Delta \mathcal{A}$, we obtain different $x_c, p_c \in \mathbb{C}$ descriptions for the wave packets and orbits, implying redundancies similar to gauge conventions. Nevertheless, the orbits following different conventions describe physics consistently, map to each other straightforwardly \cite{SuppCST}, and converge within similar steps and computational costs given the same period.

Just like in conventional semiclassical theory, we can derive quantum eigenstates, i.e., wave functions, from the orbits of the complex semiclassical theory. We achieve such goals by applying Eq. \ref{eq:coh_map2}, which maps the wave packet with ${\bf r}_c, {\bf p}_c\in \mathbb{C}$ at each instance to our familiar form with ${\bf r}_r, {\bf p}_r\in \mathbb{R}$, and incorporating the extra factor $\mathcal{A}$ upon the summation along the orbit. For instance, the resulting eigenstates of Eq. \ref{eq:ho_Ham} from benchmarks and the complex semiclassical theory compare fully consistently in Table \ref{tab:statecompare}; we have also confirmed establishing quantum eigenstates of lattice models based on the complex semiclassical theory; see Ref. \cite{SuppCST} for further results and details. We also note that the single-valuedness of such wave functions, together with the geometric phase, complex yet still in the form of $\int {\bf p_c}\cdot d{\bf r_c}$ following generalized translations $\hat W({\bf r}_c, {\bf p}_c)$, requires the discrete quantization condition in Eq. \ref{eq:QC} for a periodic orbit \cite{SuppCST}.

\emph{Lattice model examples} \textemdash The complex semiclassical theory also applies to lattice models. For example, we consider the following non-Hermitian 2D model:
\begin{equation}
\hat{H}= \sum_{\bf r} \eta c^\dagger_{{\bf r}+\hat{x}+\hat{y}}c_{\bf r} - c^\dagger_{{\bf r}+\hat{x}}c_{\bf r} - c^\dagger_{{\bf r}-\hat{x}}c_{\bf r}-V c^\dagger_{{\bf r}+\hat{y}}c_{\bf r}- V c^\dagger_{{\bf r}-\hat{y}}c_{\bf r},
\label{eq:Hamlattice}
\end{equation}
with an external magnetic field $B\hat{z}$, beyond the non-Bloch band theory \cite{Wang2018, Murakami2019, Fang2020} and previous semiclassical theory frameworks.

\begin{figure}
\includegraphics[width=0.5\linewidth]{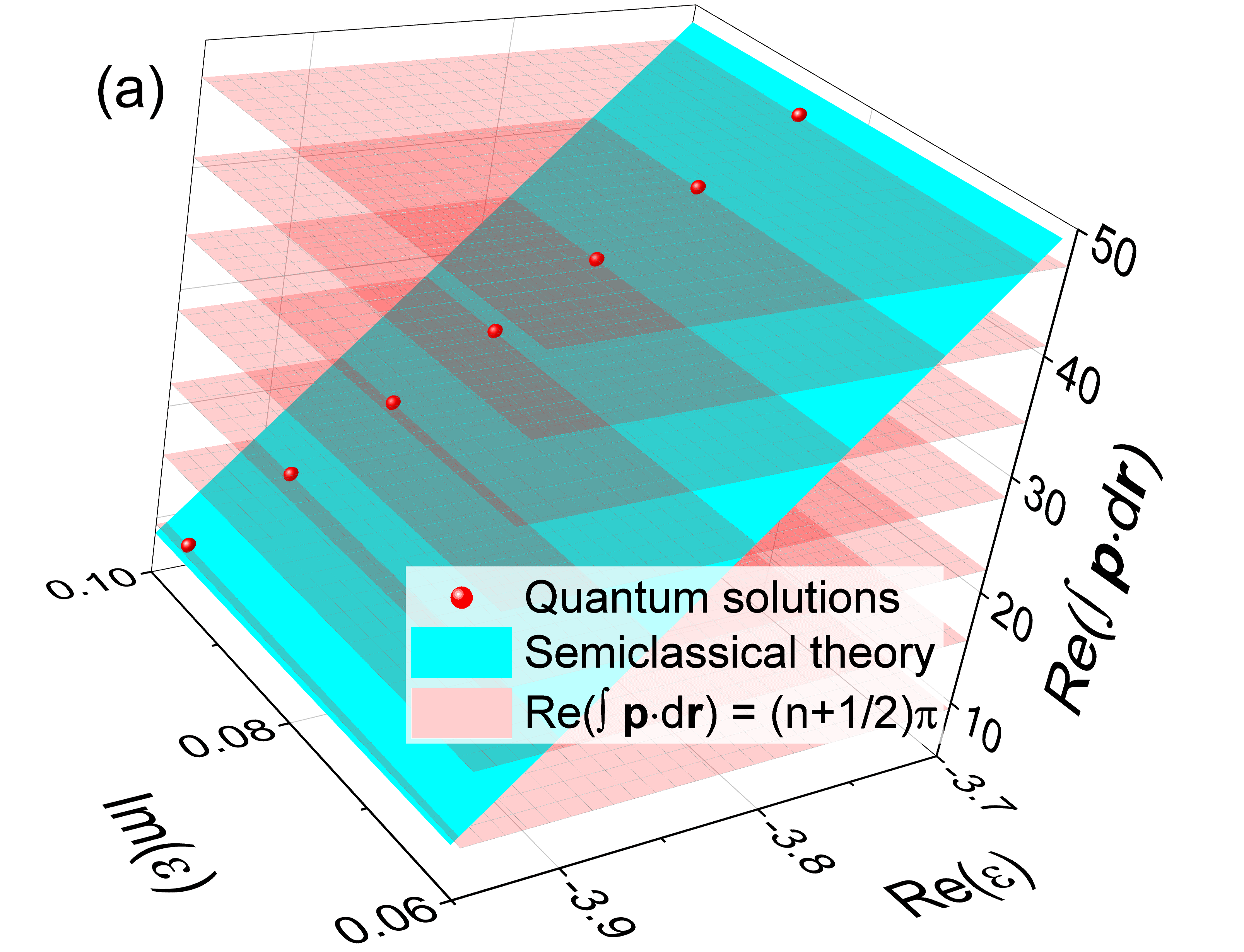}\includegraphics[width=0.5\linewidth]{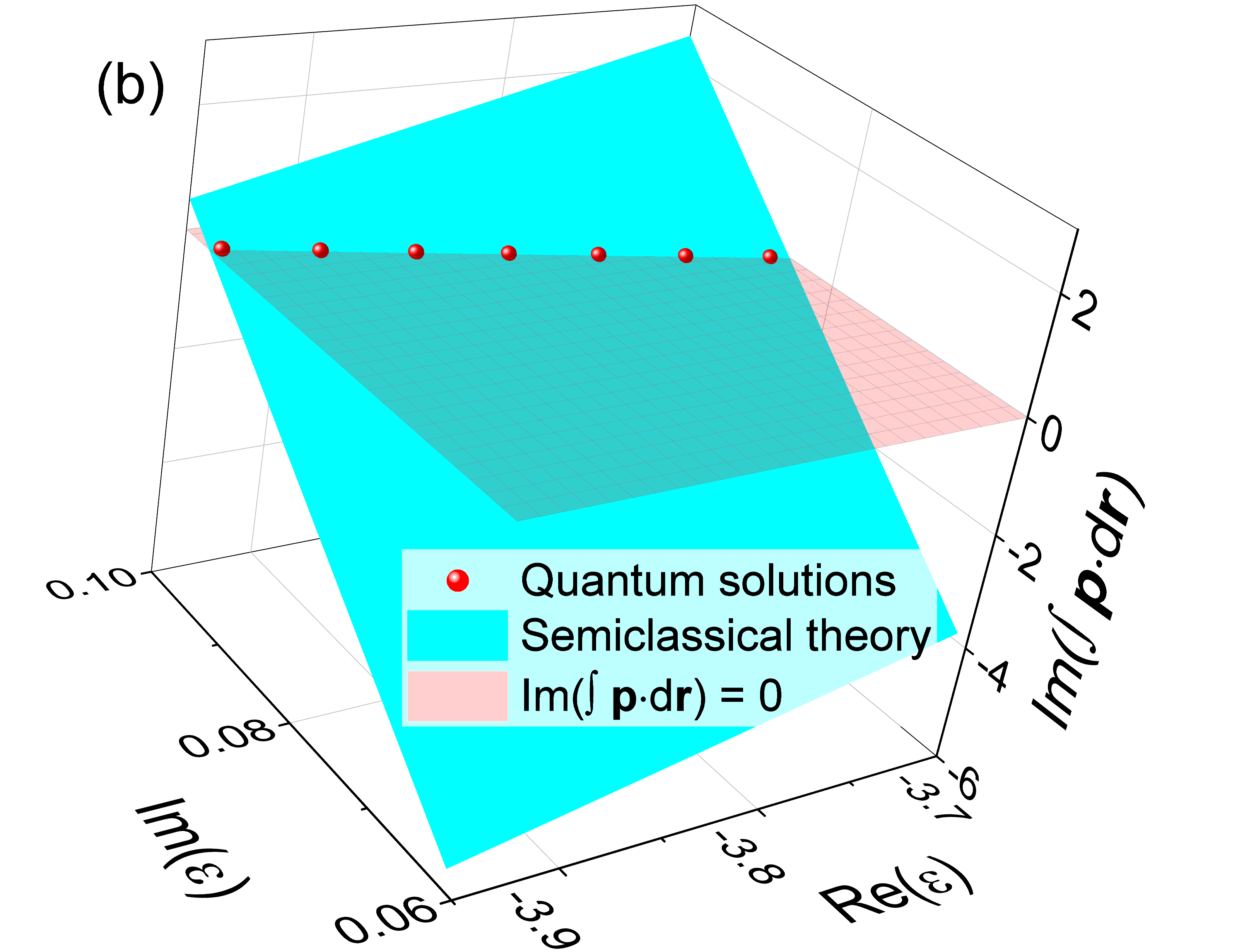}
\includegraphics[width=0.5\linewidth]{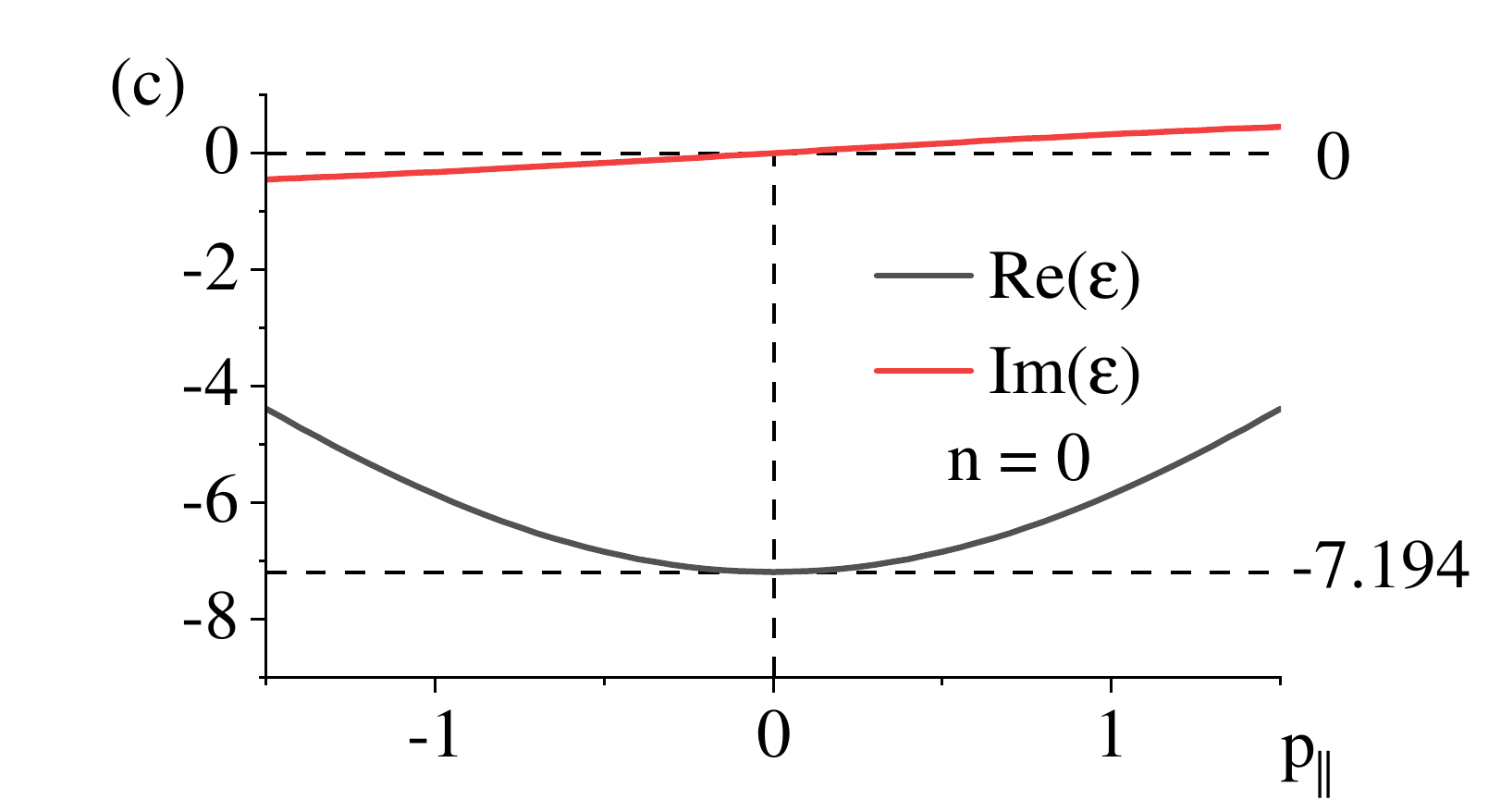}\includegraphics[width=0.5\linewidth]{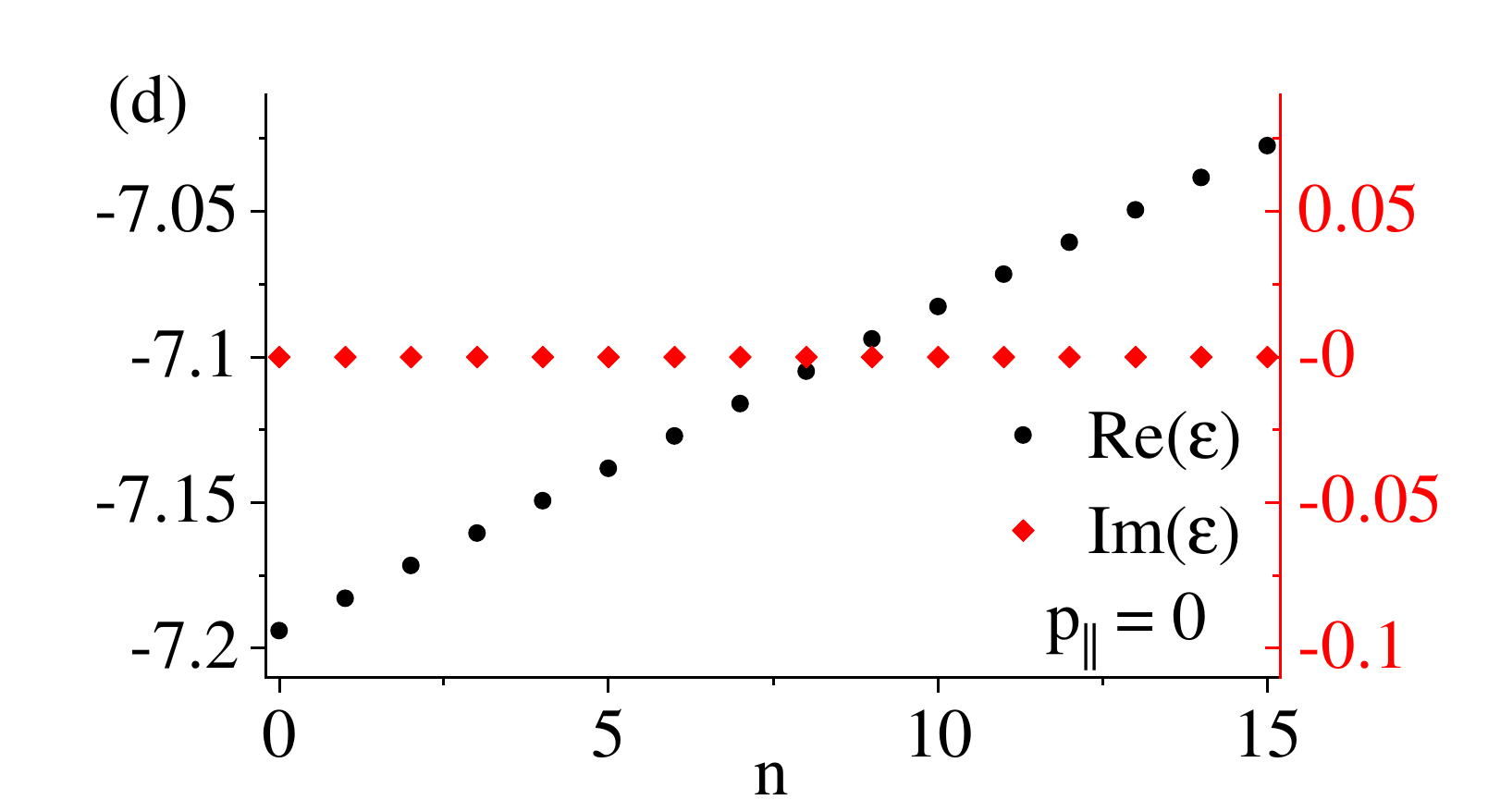}
\caption{For each complex energy $\epsilon$, we determine the satisfiability of the quantization condition through the (a) real and (b) imaginary parts of $\oint {\bf p} \cdot d{\bf r}$ over the closed orbits. The translucent planes denote $(2n+1)\pi$, $n\in \mathbb{Z}$ for the real part and $0$ for the imaginary part, which pinpoints semiclassical $\epsilon_n$ in excellent consistency with the quantum spectrum (red dots). For simpler comparisons, we have shifted the red dots vertically. $V=1.0$, $\eta = 0.1i$, and $B=0.02$. Similarly, (c) and (d) show the complex spectra of a three-dimensional lattice model with a generic magnetic field through the complex semiclassical theory \cite{SuppCST}. $p_\parallel$ is the momentum along the magnetic field. }
\label{fig:lattice_pdx}
\end{figure}

Through the complex semiclassical theory, we establish the EOMs as:
\begin{eqnarray}
\epsilon &=& -2\cos(p_x) - 2V\cos(p_y) + \eta e^{i(p_x+p_y)}, \nonumber \\
\dot{\bf p} &=& \dot{\bf r}\times B\hat{z} = \partial \epsilon / \partial {\bf p} \times B\hat{z},  \\
\dot p_x &=& B\partial \epsilon/\partial p_y = B[2V \sin(p_y) + i\eta e^{i(p_x+p_y)}], \nonumber\\
\dot p_y &=& -B \partial \epsilon/\partial p_x = -B[2\sin(p_x) + i\eta e^{i(p_x+p_y)}]. \nonumber
\label{eq:latticeEOM}
\end{eqnarray}
We track the trajectories in the complex $\bf p$ space via finite-time steps and evaluate $\oint {\bf p} \cdot d{\bf r}$ over the closed orbits, which we summarize in Figs. \ref{fig:lattice_pdx}a and \ref{fig:lattice_pdx}b. The real and imaginary parts of the quantization condition $\oint {\bf p}\cdot d{\bf r} = 2\pi(n+1/2)$ offer two constraints and thus a discrete series of energies $\epsilon_n$ on the complex $\epsilon$ plane. Such analysis generalizes straightforwardly to higher dimensions, e.g., the spectra of three-dimensional lattice models in magnetic fields as Figs. \ref{fig:lattice_pdx}c and \ref{fig:lattice_pdx}d, far beyond the capacity of previous approaches. We include further details and examples in the supplemental materials \cite{SuppCST}.

\begin{figure}
\includegraphics[width=0.98\linewidth]{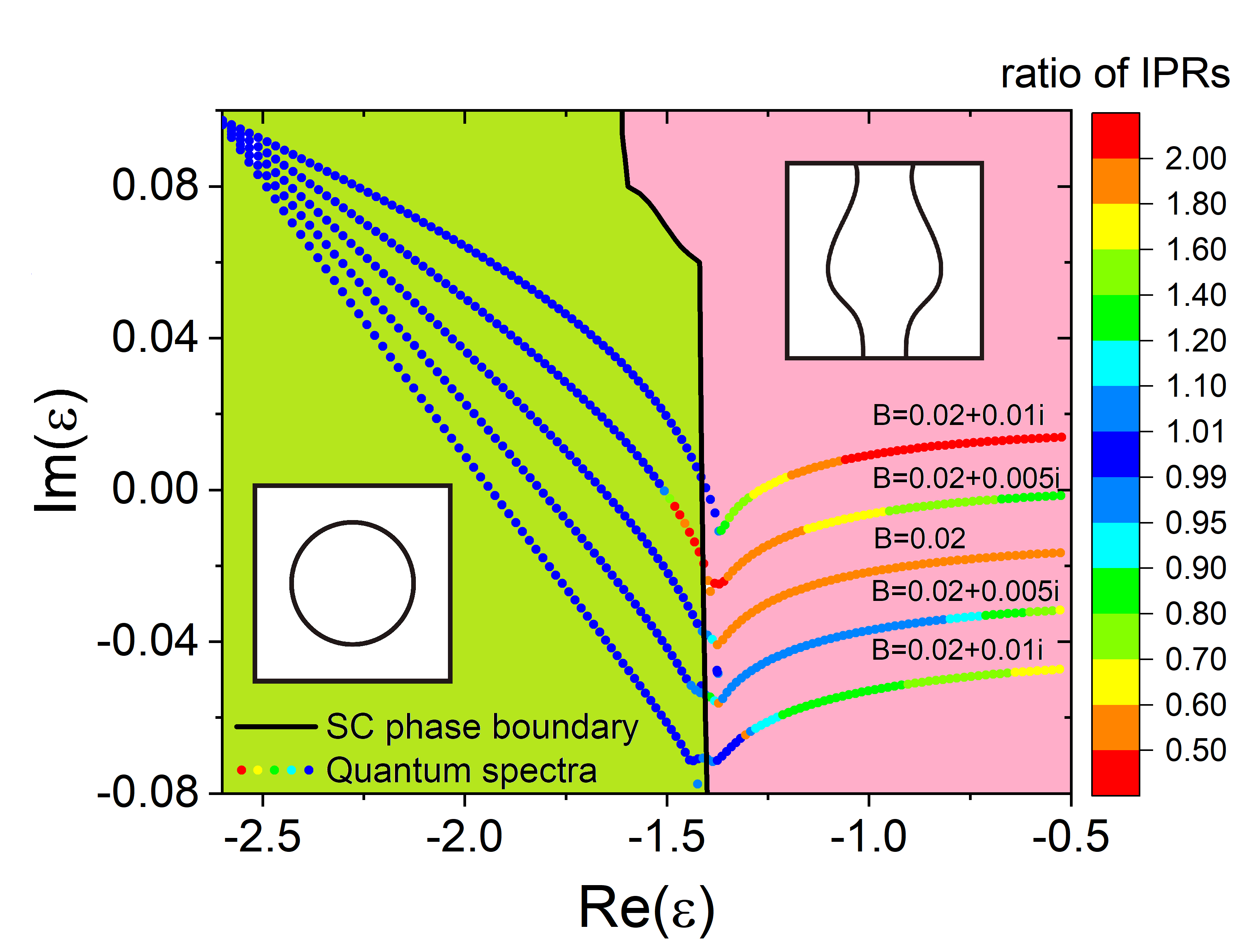}
\caption{The complex semiclassical theory characterizes the phases (bright green and pink regions) and phase boundary (black curve) for the non-Hermitian lattice model in Eq. \ref{eq:Hamlattice} based on the orbits' connectivity. The dots and their color scales denote the quantum energy spectrum $\epsilon_n$ and each eigenstate's ratio of IPRs upon doubling the system size - a measure of localization. The insets illustrate the two phases' schematic orbits. $V=0.3$, $\eta=0.1i$. }
\label{fig:lattice_pd}
\end{figure}

The complex semiclassical theory also offers straightforward recipes for analyzing phases and transitions in non-Hermitian quantum systems. For example, the interplay between the quantum Hall effect and the NHSE has received much recent attention \cite{XingPRB2022, LuPRL2021}. The complex semiclassical theory separates these two phases by a Lifshitz transition: the former cases possess closed orbits $p(T) = p(0)$, thus accompanied by discrete Landau levels and localized quantum Hall states; in contrast, the latter case instead obtains orbits satisfying ${\bf p}(T) = {\bf p}(0)+ {\bf G}$, where ${\bf G} \neq 0$ is a reciprocal lattice vector. Accordingly, we summarize the phases of the non-Hermitian lattice model in Eq. \ref{eq:Hamlattice} with $V=0.3$ in Fig. \ref{fig:lattice_pd}, which indicates a mobility edge in the complex $\epsilon$ space with ${\bf G}=2\pi \hat y$ on the right-hand side. Consequently, the wave packet receives a nontrivial displacement within each period: $\Delta{\bf r} = -\oint d{\bf p} \times \hat z / B = - 2\pi \hat x /B$, a clear signature of delocalization along the $\hat x$ direction and paving the road for the NHSE - unique insights on transport and localization properties from the complex semiclassical theory.

For comparison, we calculate the quantum eigenstates' inverse participation ratio (IPR): $(\sum_x |\psi(x)|^4)^{-1}$ \cite{Denner2021}. In and only in the quantum Hall phase, the IPRs remain robust upon enlarging the system and signal localization, consistent with the complex semiclassical theory. Sharp kinks also appear in the complex spectra, implicating transitions near the semiclassically predicted boundary. We note that the diagonalization of non-Hermitian Hamiltonians suffers accumulated error \cite{Fanghu2020, Zhong2021amplification, FuPRL2018}, especially for delocalized states on large systems, which caused the IPRs' instability on the right-hand side in Fig. \ref{fig:lattice_pd}. In comparison, the complex semiclassical theory excels at small magnetic fields, where the magnetic unit cells are large. Further, with a soft potential $V(x)$ characterizing boundaries, the NHSE emerges naturally as a position dependent $\mathcal{A}$ factor from the complex semiclassical theory, demonstrated in the supplemental materials \cite{SuppCST}.

\emph{Discussions} \textemdash We have established a complex semiclassical theory more generally applicable to non-Hermitian quantum systems with an analytical continuation of physical variables. We have also provided a strategy to locate closed orbits and eigenstates. We demonstrate our framework's applicability, accuracy, and efficiency on various continuous and lattice models, generally unattainable by previous ansatzes due to system sizes, dimensions, potentials, magnetic fields, and complex spectra.

There are several interesting directions for further generalizations. We discuss the introduction of multiple bands and the Berry curvature ${\bf \Omega}$ in Ref. \cite{SuppCST}. On the other hand, we may consider the analytical continuation of the remaining two physical quantities in the EOMs, the Berry curvature \cite{IlanPRB2020, Fan_2020} and the magnetic field, to the complex domains. Indeed, the physics of the model in Eq. \ref{eq:Hamlattice} in a complex-valued magnetic field $\bf B$ can be satisfactorily captured by the complex semiclassical theory (Fig. \ref{fig:lattice_pd} and Ref. \cite{SuppCST}). The complex semiclassical theory may also apply beyond the framework of non-Hermitian Hamiltonians. For example, it is straightforward to incorporate dissipative friction, such as ${\bf F_\mu}= -\mu{ \bf p}$, into the EOMs, which forbids closed orbits for $t\in \mathbb{R}$; the complex semiclassical theory opens up the possibility of $t\in \mathbb{C}$, thus potential approaches and insights for dissipative or driven quantum systems \cite{SuppCST}.

\emph{Acknowledgments} \textemdash We acknowledge helpful discussions with Ryuichi Shindou, Junren Shi, and Haoshu Li. We also acknowledge support from the National Key R\&D Program of China (No.2022YFA1403700) and the National Natural Science Foundation of China (No.12174008 \& No.92270102).

\bibliography{ref}

\newpage

\appendix

\section{I. The equation of motion for ${\bf p}_c$}

In the main text, we have established the semiclassical equation of motion (EOM) for the complex variable ${\bf r}_c$; in this section, we derive the second line of Eq. 1 in the main text - the EOM for ${\bf p}_c$. Similar to the main text, we begin with the evolution of the wave packet $|{\bf r}_c(t+\tau), {\bf p}_c(t+\tau)\rangle = e^{-i\hat{H}\tau}|{\bf r}_c(t), {\bf p}_c(t)\rangle$ after a short time step $\tau \rightarrow 0$:
\begin{eqnarray}
\dot{\bf p}_c(t) &=& \left[{\bf p}_c(t+\tau) - {\bf p}_c(t)\right]/\tau \label{eq:TR_deri} \nonumber\\
&=& \langle {\bf r}_c(t), {\bf p}_c(t)| e^{i\hat H \tau} \hat{\bf p} e^{-i\hat{H}\tau} - \hat{\bf p}|{\bf r}_c(t), {\bf p}_c(t) \rangle / \tau \\
&=& -i\langle 0| \hat W(-{\bf r}_c(t), -{\bf p}_c(t)) \left[\hat{\bf p},  \hat{H}\right] \hat W({\bf r}_c(t), {\bf p}_c(t))|0\rangle \nonumber \\
&=& \langle 0|\left(-\frac{\partial \hat H}{\partial \hat{\bf r}}+\frac{\partial \hat H}{\partial \hat{\bf p}}\times {\bf B}\right)(\hat{\bf r}+{\bf r}_c(t), \hat{\bf p}+{\bf p}_c(t))|0\rangle, \nonumber
\end{eqnarray}
where for the last line, we note that in addition to $[r_i, p_j]=i\delta_{ij}$, we have $[p_i, p_j]=i\epsilon_{ijk}B_k$ in the presence of a magnetic field. For reasons we have stated in the main text, we only need to keep the $m=n=0$ term in the Taylor expansions:
\begin{eqnarray}
\frac{\partial \hat H}{\partial \hat{\bf r}}(\hat{\bf r}+{\bf r}_c(t), \hat{\bf p}+{\bf p}_c(t)) &=& \sum_{mn}\frac{\hat{\bf r}^n \hat{\bf p}^m}{n!m!}\frac{\partial \hat{H}}{\partial^{n+1} \hat{\bf r} \partial^{m} \hat{\bf p}}({\bf r}_c(t), {\bf p}_c(t)), \nonumber\\
\frac{\partial \hat H}{\partial \hat{\bf p}}(\hat{\bf r}+{\bf r}_c(t), \hat{\bf p}+{\bf p}_c(t)) &=& \sum_{mn}\frac{\hat{\bf r}^n \hat{\bf p}^m}{n!m!}\frac{\partial \hat{H}}{\partial^n \hat{\bf r} \partial^{m+1} \hat{\bf p}}({\bf r}_c(t), {\bf p}_c(t)), \nonumber\\
\end{eqnarray}
equivalent to replacing operators $\hat{\bf r}$ and $\hat{\bf p}$ in $\partial\hat{H}/\partial{\bf r}$ and $\partial\hat{H}/\partial{\bf p}$ with the complex variables ${\bf r}_c$ and ${\bf p}_c$:
\begin{eqnarray}
    \dot{\bf p}_c(t) &\approx& -\frac{\partial \epsilon({\bf p}_c, {\bf r}_c)}{\partial {\bf r}_c} + \frac{\partial \epsilon({\bf p}_c, {\bf r}_c)}{\partial {\bf p}_c} \times {\bf B} \nonumber\\
    &\approx& {\bf E} + \dot{\bf r}_c(t) \times {\bf B}.
\end{eqnarray}

\section{II. Continuous model examples}

\subsection{IIA. Closed orbits of non-Hermitian harmonic oscillators from a quantum perspective}

The closed orbits in Eq. 5 in the main text are solvable through the EOMs, which are linear different equations in the current case. In this subsection, we study the quantum evolution of the wave packets under the Hamiltonian in Eq. 3 in the main text, which offers an alternative perspective of the orbits. As shown in the main text, we can re-express the Hamiltonian as:
\begin{eqnarray}
\hat{H} &=& \omega \left(b^\dagger a + 1/2\right), \nonumber\\
\hat a &=& \frac{2\alpha\hat x+(\eta+i\omega)\hat p}{2\sqrt{\alpha\omega}}, \nonumber\\
\hat b^\dagger &=& \frac{2\alpha\hat x+(\eta-i\omega)\hat p}{2\sqrt{\alpha\omega}},
\end{eqnarray}
where $\omega = \sqrt{4\alpha\beta-\eta^2}$ is generally complex. Although $\hat b^\dagger \neq  \hat a^\dagger$, we can still make use of the commutation relation $[\hat a, \hat{b}^\dagger]=1$ to obtain the ladder algebra:
\begin{eqnarray}
\hat b^\dagger |n\rangle &=& \sqrt{n+1}|n+1\rangle, \nonumber\\
\hat a |n\rangle &=& \sqrt{n}|n-1\rangle, \nonumber\\
\hat b^\dagger \hat a |n\rangle &=& n |n\rangle, \nonumber
\end{eqnarray}
where $\hat a|0\rangle = 0$. Therefore, $|n\rangle$ and $\epsilon_n = (n+1/2)\omega$ are the (right) eigenstates and energy eigenvalues of $\hat{H}$, respectively, $n\in \mathbb{Z}$.

Now, we start from a wave-packet state $|z_0\rangle = e^{-|z_0|^2/2}e^{z_0 \hat b^\dagger}|0\rangle$, $z_0 \in \mathbb{C}$, whose evolution under $\hat{H}$ is $e^{-i\omega t/2} |z_t\rangle$, where $z_t=e^{-i\omega t} z_0$. Consequently, the wave packet's center of mass (COM) exhibits the following time dependence:
\begin{eqnarray}
    \langle \hat x\rangle(t) &=& \langle 0|e^{-z_t \hat b^\dagger} e^{|z_t|^2/2} \hat{x} e^{-|z_t|^2/2} e^{z_t \hat b^\dagger}|0\rangle \nonumber\\
    &=& \langle 0|\hat{x}- z_t [\hat b^\dagger, \hat x]|0\rangle  \nonumber\\
    &=& \frac{(\omega+i\eta)z_t}{2\sqrt{\alpha\omega}}= z_2 e^{-i\omega t}, \\
    \langle \hat p\rangle(t) &=& \langle 0|e^{-z_t \hat b^\dagger} e^{|z_t|^2/2} \hat{p} e^{-|z_t|^2/2} e^{z_t \hat b^\dagger}|0\rangle \nonumber\\
    &=& \langle 0|\hat{p}- z_t [\hat b^\dagger, \hat p]|0\rangle  \nonumber\\
    &=& \frac{-2i\alpha z_t}{2\sqrt{\alpha\omega}}= -\frac{(i\omega +\eta)}{2\beta} z_2 e^{-i\omega t}, \nonumber
\end{eqnarray}
where we have set $z_2 = (\omega+i\eta)z_0/2\sqrt{\alpha\omega}$. We have used the facts that $\langle 0|\hat{x}|0\rangle = 0 $ and $\langle 0|\hat{p}|0\rangle = 0$, guaranteed by parity symmetry, or more generally, the COM definition of wave packet $|0\rangle$. These results are consistent with the $z_2$ terms in Eq. 5 in the main text obtained via semiclassical equations of motion (EOMs). Note that we have employed the biorthogonal basis, thus inverse instead of Hermitian conjugate, in the expectation values $\langle\hat{x}\rangle(t)$ and $\langle\hat{p}\rangle(t)$, which can take on complex values.

To obtain the $z_1$ terms, on the other hand, we may start with a different basis of ladder states: $\hat b^\dagger |0\rangle' = 0$, $\hat a|n\rangle' = \sqrt{n+1} |n+1\rangle'$. The corresponding wave-packet state evolves with an $e^{i\omega t}$ factor independently from the $z_2$ terms.

\subsection{IIB. Continuous model examples with higher-order terms}

In this subsection, we examine additional continuous models - first, a non-Hermitian quantum model with higher-order $\hat{x}$:
\begin{equation}
\hat{H}= \beta \hat{x}^4 +\gamma \hat{p}^2 + \alpha \hat{a}^2,
\label{eq:spmodel1}
\end{equation}
where $\alpha, \beta, \gamma \in \mathbb{C}$. With $\hat{x}=\frac{1}{\sqrt{2}}\left(\hat{a}+\hat{a}^\dagger\right)$ and $\hat{p}=\frac{i}{\sqrt{2}}\left(\hat{a}^\dagger-\hat{a}\right)$ in terms of the ladder operators, we can re-express the Hamiltonian as:
\begin{eqnarray}
    \hat{H}&=&\frac{\beta}{4}[\hat{a}^4+\left( \hat{a}^\dagger \right)^4+4\hat{a}^\dagger \hat{a}^3+ 4(\hat{a}^\dagger)^3  \hat{a}  +6(\hat{a}^\dagger)^2 + 6\hat{a}^2 \nonumber\\
    &+& 6n^2+6n+3] -\frac{\gamma}{2}[(\hat{a}^\dagger)^2 +\hat{a}^2-2n-1]+\alpha\hat{a}^2,
\end{eqnarray}
which we diagonalize in the occupation number basis $\hat{n}|n\rangle = n|n\rangle$ numerically (with a truncation at sufficiently large $n$) for benchmark spectra.

\begin{figure}
\includegraphics[width=0.49\linewidth]{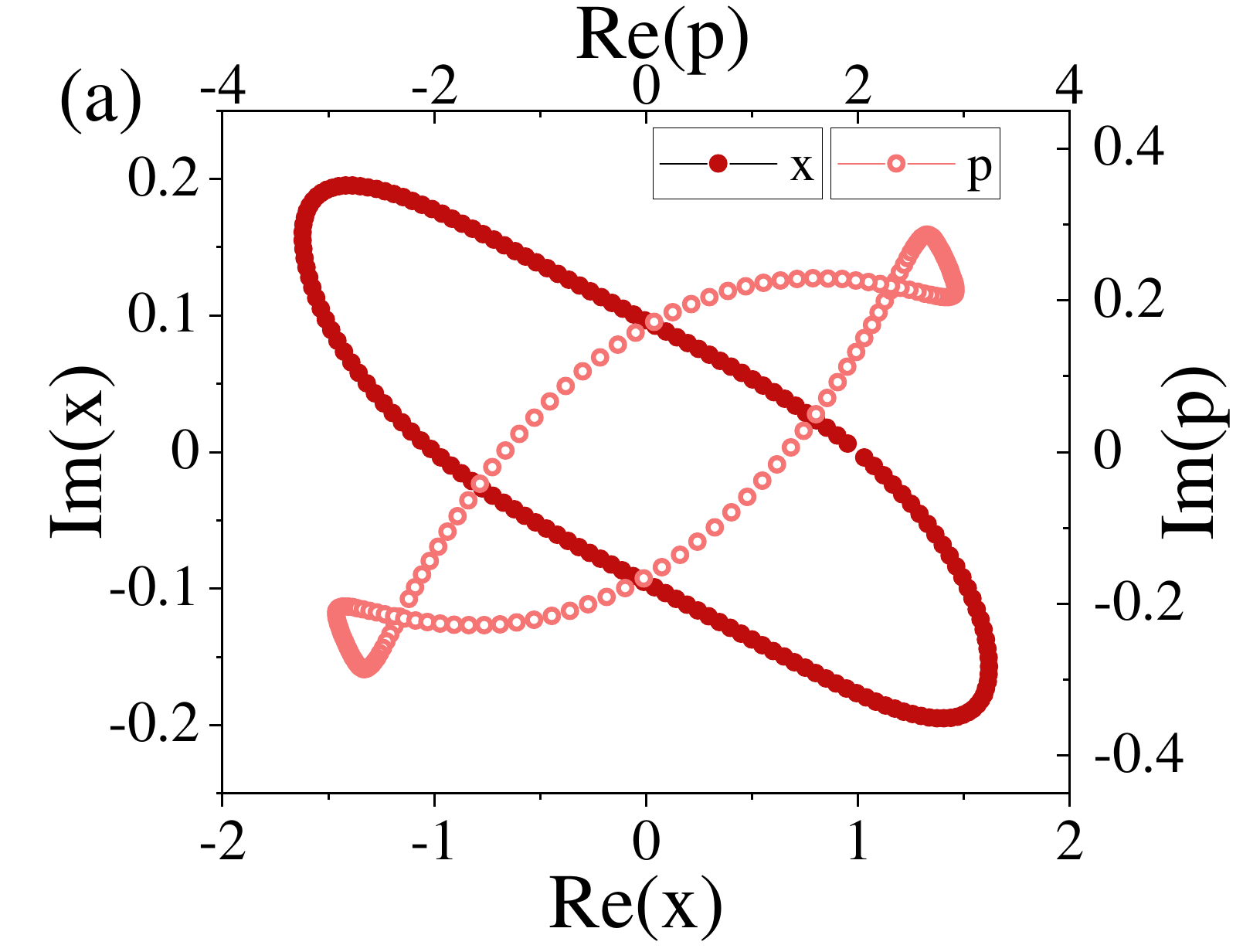}
\includegraphics[width=0.49\linewidth]{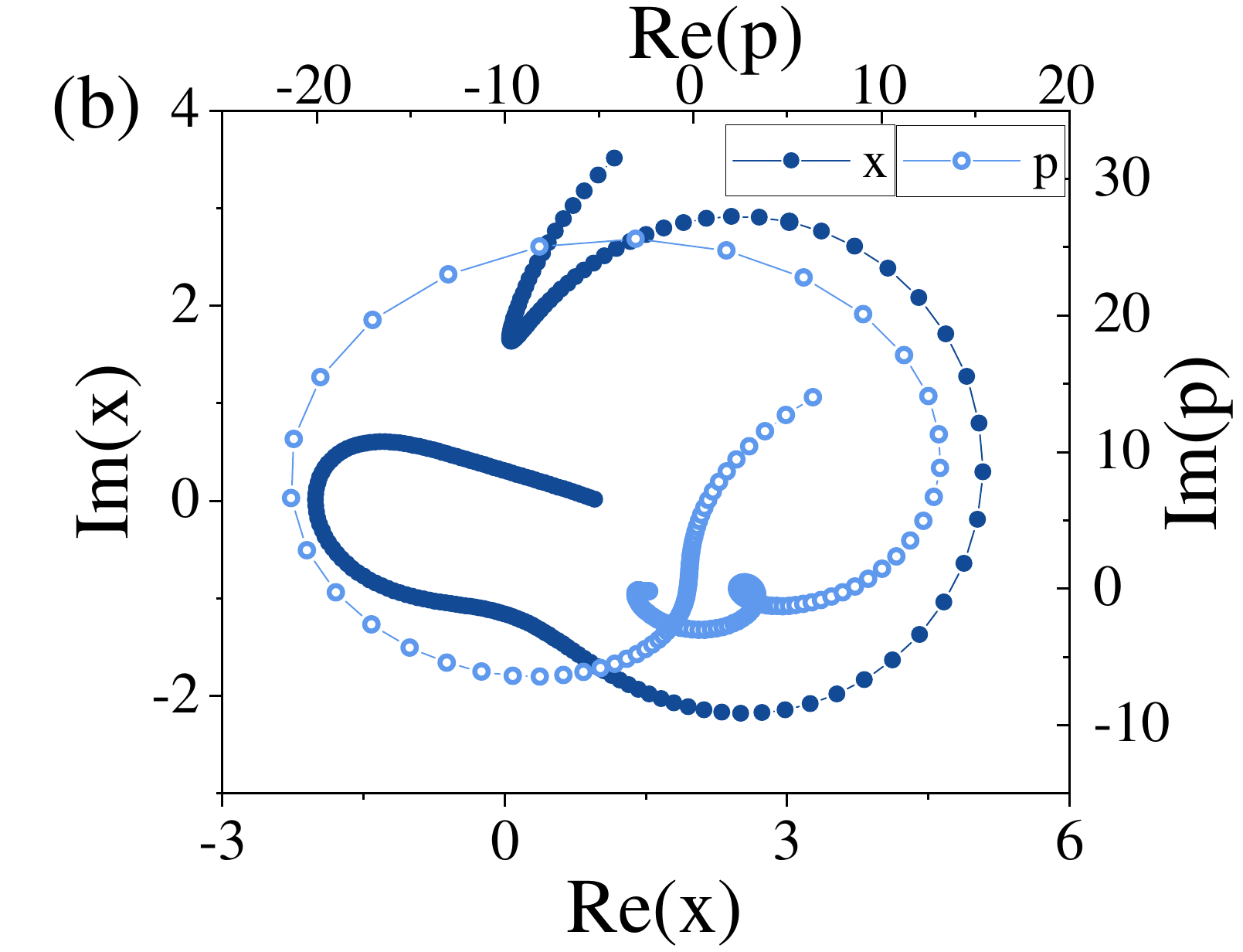}
\includegraphics[width=0.49\linewidth]{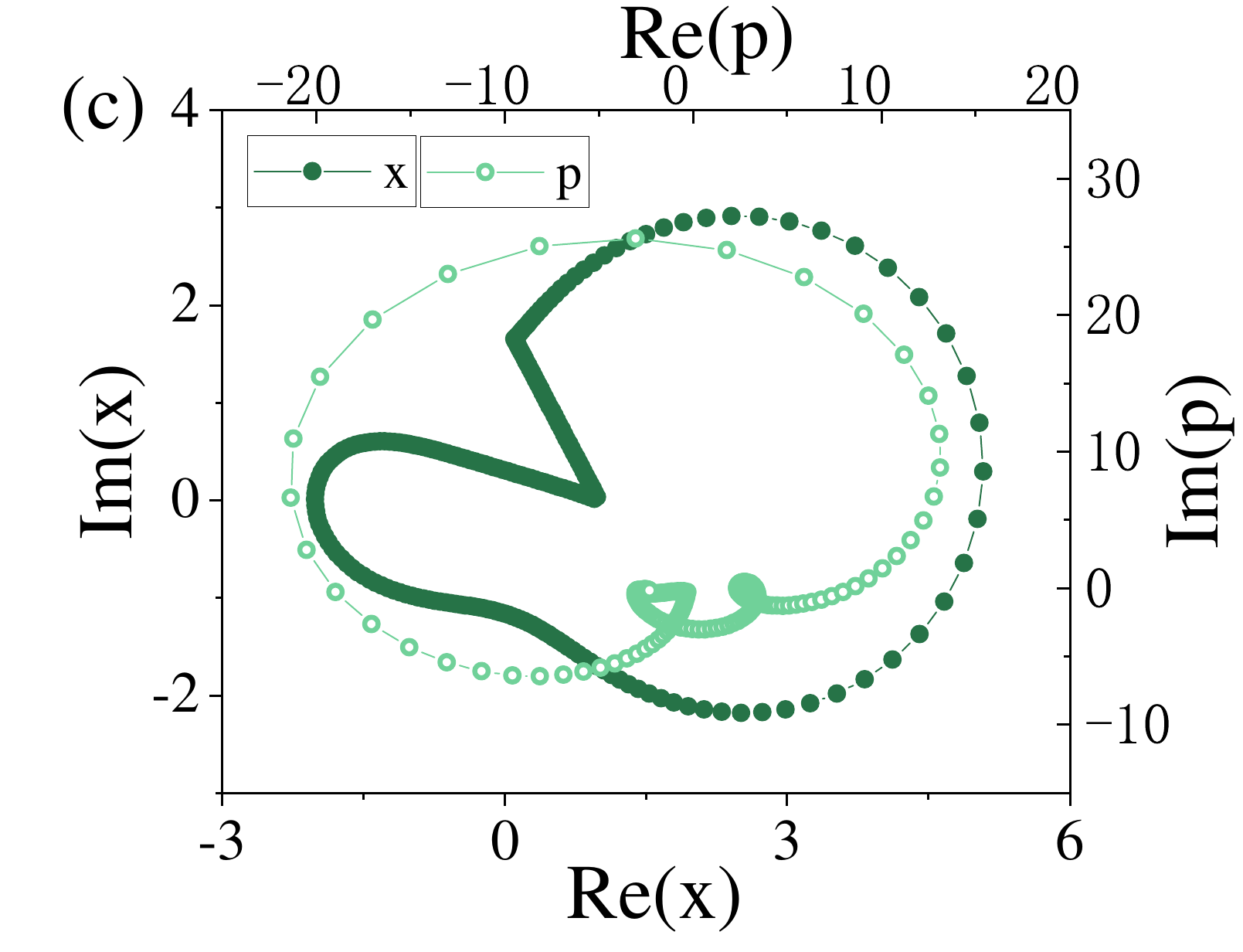}
\includegraphics[width=0.49\linewidth]{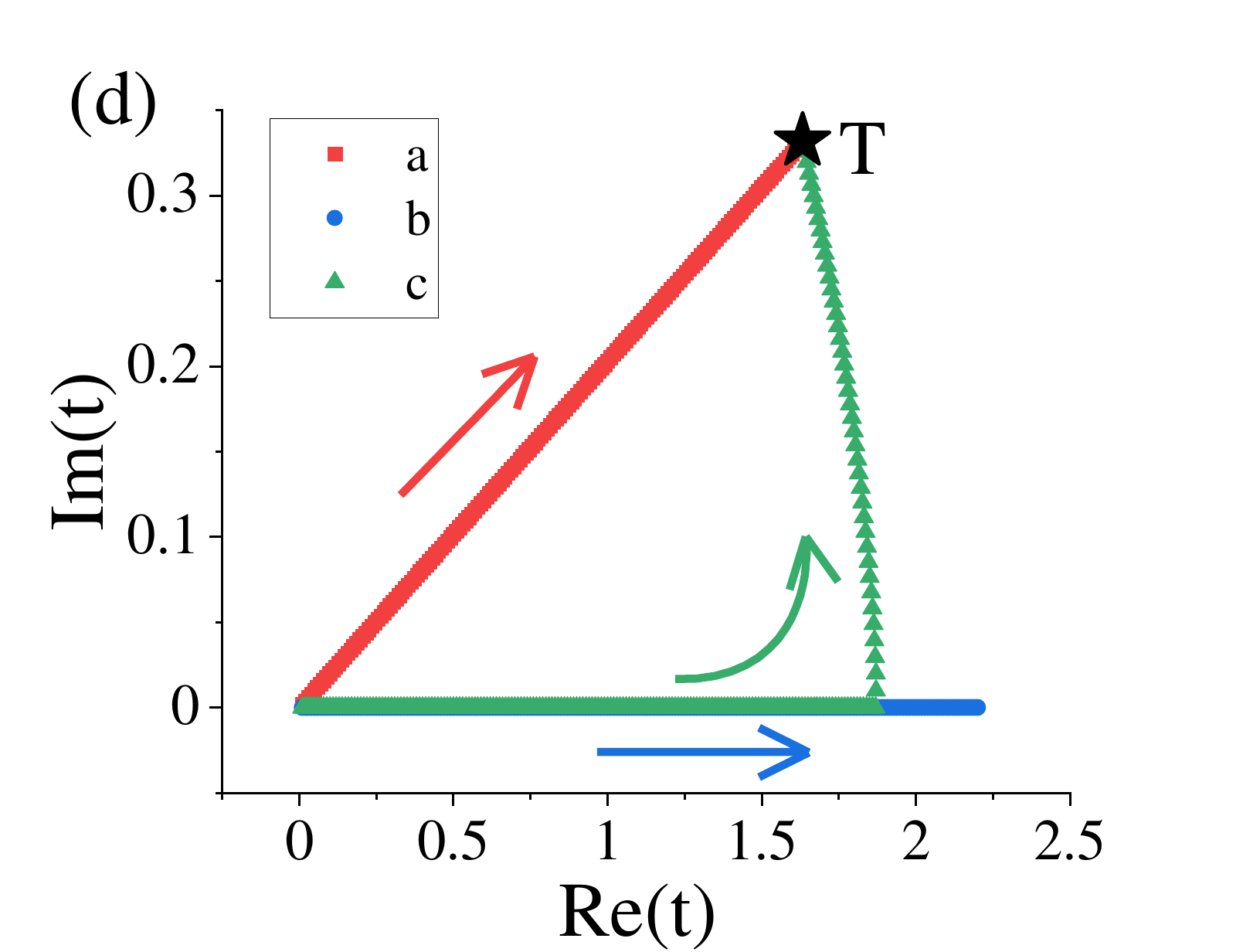}
\caption{Following the EOMs, the semiclassical trajectories of the model in Eq. \ref{eq:spmodel1} form (a) closed loops along the $t/T \in \mathbb{R}$ direction and end up in (b) open paths along the $t \in \mathbb{R}$ direction. Without knowing $T$, we may still obtain (c) closed loops by strategically adjusting $dt$. (d) The corresponding trajectories in the complex $t$ plane show the equivalence between different closed orbits reaching the complex period $T$. $\alpha = 0.3+0.5i$, $\beta=1.0+0.1i$, $\gamma = 1.0-0.1i$, and $\oint p\cdot dx = 5\pi$ indicating the complex energy $\epsilon=7.04+1.24i$ qualifies for a quantized energy level. }
\label{fig:spmodel1}
\end{figure}

Using complex semiclassical theory, we begin with the energy relation $\epsilon = \beta x^4 + \gamma p^2 + \alpha(x+ip)^2/2$. The resulting EOMs are:
\begin{eqnarray}
\dot x &=& i\alpha x  + \left(2\gamma-\alpha\right) p,  \nonumber\\
\dot p &=& -4\beta x^3-\alpha x - i\alpha p,
\label{eq:qdr_EOM}
\end{eqnarray}
which we employ to search for the closed orbits. The strategy of $dt$ phase adjustment we mentioned in the main text is especially helpful in the process; see illustrations in Fig. \ref{fig:spmodel1}. In particular, the complex energies with qualified quantization conditions are consistent with the quantum benchmarks; see examples in Table \ref{tab:spmodel1}.

\begin{table}[]
    \centering
    \renewcommand{\arraystretch}{1.2}
    \setlength\tabcolsep{1.4mm}{
    \begin{tabular}{|c|c|c|c|c|} \hline
         $n$  & 1 & 2 & 5 & 6.39+0.89i  \\ \hline
         $Re[\epsilon_{n}]$  & 3.622 & 7.037 & 19.861 & 10\\ \hline
         $Im[\epsilon_{n}]$  & -0.558 & -1.240 & -3.936 & 0\\ \hline
         $Re[\oint p \cdot dx]$ & 9.518 & 15.774 & 34.588 & 20.075 \\ \hline
         $Im[\oint p \cdot dx]$ & 0.005 & -0.001 & 0.001 & 2.781 \\ \hline
         Quantum & Yes & Yes & Yes & No \\ \hline
    \end{tabular}}
    \caption{The real and imaginary parts of the energies and geometric phases of closed orbits following the complex semiclassical EOMs are consistent with the quantization condition $\oint p \cdot dx=(2n+1)\pi$ for quantized, complex energy eigenvalues for the model in Eq. \ref{eq:spmodel1}. The parameters are $\alpha = 0.3+0.5i$, $\beta=1.0+0.1i$, and $\gamma = 1.0-0.1i$.}
    \label{tab:spmodel1}
\end{table}

More generally, the complex semiclassical theory applies to non-Hermitian quantum systems with potential and kinetic terms beyond quadratic order. For instance, we consider the following continuous model:
\begin{equation}
\hat{H}=\beta \hat{p}^4 + \gamma \hat{x}^4 + \alpha \hat{a}^2,
\label{eqmodel3}
\end{equation}
where $\alpha, \beta, \gamma \in \mathbb{C}$. In terms of the ladder operators, the Hamiltonian becomes:
\begin{eqnarray}
    \hat{H}&=&\frac{\beta}{4}[\hat{a}^4+\left( \hat{a}^\dagger \right)^4+4\hat{a}^\dagger \hat{a}^3 + 4(\hat{a}^\dagger)^3\hat{a}  +6(\hat{a}^\dagger)^2 + 6\hat{a}^2 \nonumber\\
    &+& 6n^2+6n+3] + \frac{\gamma}{4}[\hat{a}^4+\left( \hat{a}^\dagger \right)^4- 4\hat{a}^\dagger \hat{a}^3 - 4(\hat{a}^\dagger)^3 \hat{a} \nonumber\\
    &-&6(\hat{a}^\dagger)^2 - 6\hat{a}^2 + 6n^2+6n+3]+\alpha\hat{a}^2,
\end{eqnarray}
which we also diagonalize numerically for the benchmark spectra.

\begin{figure}
\includegraphics[width=0.49\linewidth]{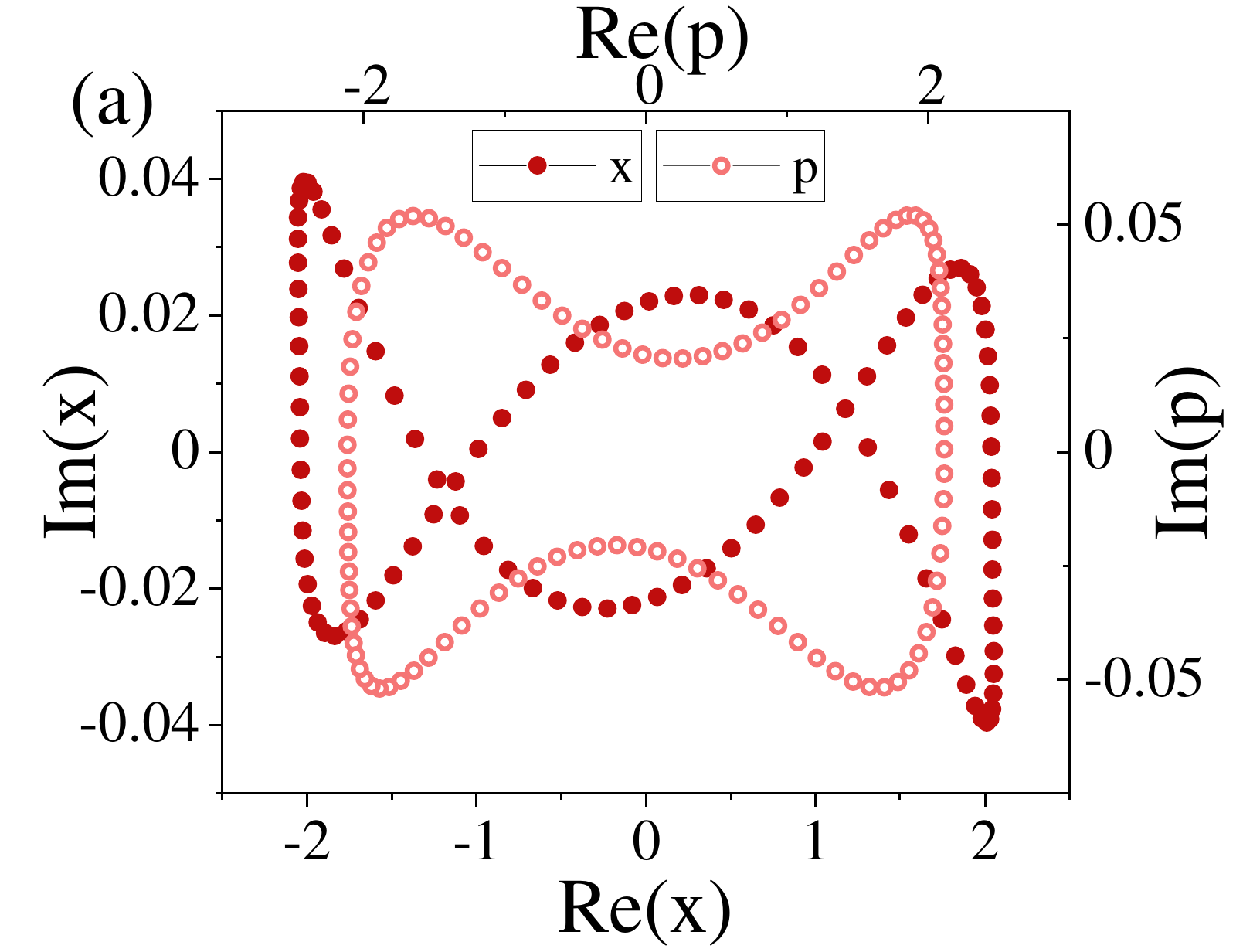}
\includegraphics[width=0.49\linewidth]{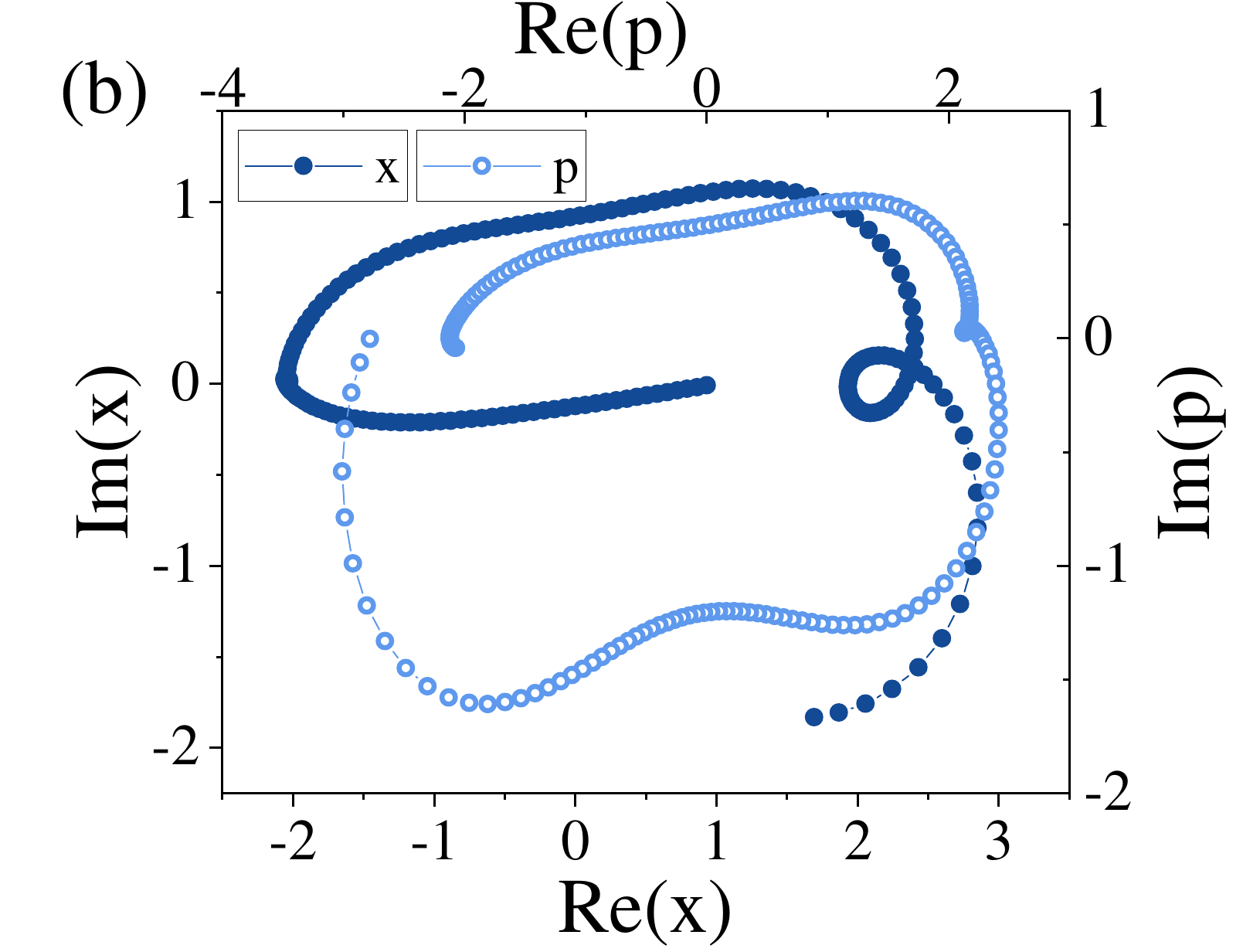}
\includegraphics[width=0.49\linewidth]{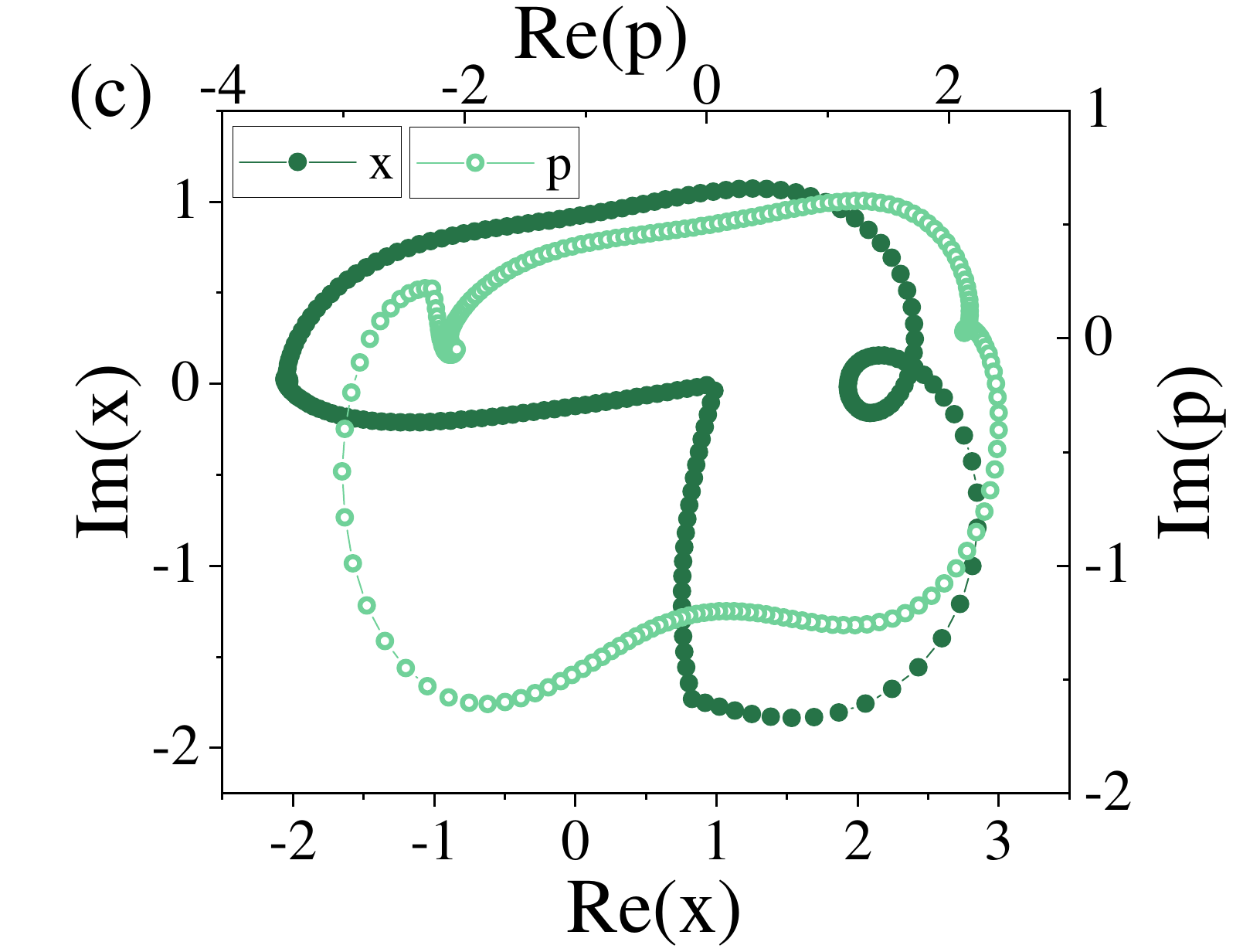}
\includegraphics[width=0.49\linewidth]{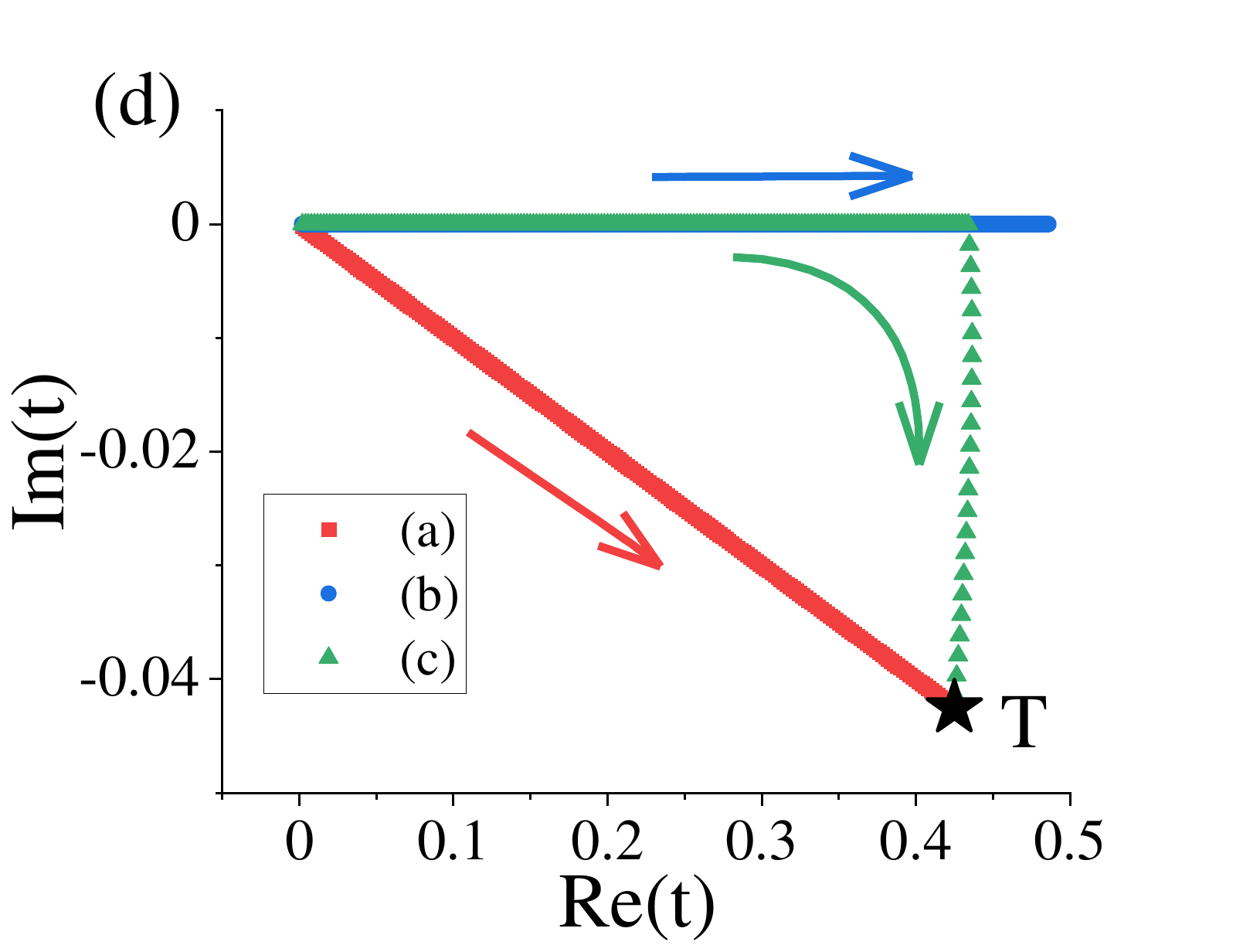}
\caption{Following the EOMs, the semiclassical trajectories of the quartic model in Eq. \ref{eqmodel3} form (a) closed loops along the $t/T \in \mathbb{R}$ direction and end up in (b) open paths along the $t \in \mathbb{R}$ direction. Without knowing $T$, we may still obtain (c) closed loops by strategically adjusting $dt$. (d) The corresponding trajectories in the complex $t$ plane show the equivalence between different closed orbits reaching the complex period $T$. $\alpha = 0.5+0.3i$, $\beta=1.0+0.1i$, and $\gamma = 1.0+0.1i$. The trajectories' energy is $\epsilon=18.65+1.87i$ with a geometric phase $\oint p\cdot dx =16.0101+0.0002i \approx 5\pi$. }
\label{fig:spmodel2}
\end{figure}

Semiclassically, the EOMs following energy relation are:
\begin{eqnarray}
\dot x &=& 4\gamma p^3 - \alpha p + i\alpha x,  \nonumber\\
\dot p &=& -4\beta x^3- \alpha x - i\alpha p,
\label{eq:qdr_EOM}
\end{eqnarray}
which we use to establish closed orbits (Fig. \ref{fig:spmodel2}). We also evaluate the integral $\oint p\cdot dx$ over the closed orbits, which determines their quantization conditions. We demonstrate several examples in Table \ref{tab:spmodel2}. Interestingly, though relatively small, there are noticeable departures of the qualified orbits' $\oint p\cdot dx$ from the exact half-integer values of $2\pi$, especially at small $n$: 9.903 versus 9.425 for $n=1$, 16.010 versus 15.708 for $n=2$. Such differences contrast sharply with previous examples of continuous models, where we have at least four or five digits of accuracy. Physically, the model in Eq. \ref{eqmodel3} deviates drastically from the quadratic form; thus, the semiclassical approximation suffers at small $n$, where the wave packets' scale and shape become relatively important and non-negligible. Indeed, such deviations vanish asymptotically at larger $n$, e.g., $n \ge 5$.

\begin{table}[]
    \centering
    \renewcommand{\arraystretch}{1.2}
    \setlength\tabcolsep{1.4mm}{
    \begin{tabular}{|c|c|c|c|c|c|} \hline
         $n$  & 1 & 2 & 5 & 10 & 3.72-0.19i  \\ \hline
         $Re[\epsilon_{n}]$  & 7.137 & 18.645 & 87.540 & 317.226 & 10\\ \hline
         $Im[\epsilon_{n}]$  & 0.721 & 1.871 & 8.760 & 31.729 & 0\\ \hline
         $Re[\oint p \cdot dx]$ & 9.903 & 16.010 & 34.694 & 66.047 & 11.680 \\ \hline
         $Im[\oint p \cdot dx]$ & 0.001 & 0.000 & 1e-04 & 3e-05 & -0.586 \\ \hline
         Quantum & Yes & Yes & Yes & Yes & No \\ \hline
    \end{tabular}}
    \caption{The real and imaginary parts of the energies and geometric phases of closed orbits following the complex semiclassical EOMs are nearly consistent with the quantization condition $\oint p \cdot dx=(2n+1)\pi$ for quantized, complex energy eigenvalues for the model in Eq. \ref{eqmodel3}, especially for relatively larger values of $n$. The parameters are $\alpha = 0.5+0.3i$, $\beta=1.0+0.1i$, and $\gamma = 1.0+0.1i$. }
    \label{tab:spmodel2}
\end{table}

These results show that within the limits of validity, our complex semiclassical theory generally applies to continuous models of non-Hermitian quantum systems.

\section{III. Lattice model examples}

\subsection{IIIA. Complex spectra of a three-dimensional non-Hermitian lattice model in generic magnetic field}

In this subsection, we use the complex semiclassical theory to showcase the straightforward derivation of complex spectra, e.g., Figs. 2c and 2d in the main text, for a non-Hermitian lattice model with a generic magnetic field in three spatial dimensions, which are beyond the reach of previous approaches.

In particular, we consider the following non-Hermitian 3D lattice model:
\begin{eqnarray}
\hat H &=& - t' \sum_{\bf r, \delta} c^\dagger_{\bf r+\hat \delta}  c_{\bf r} -\sum_{\bf r} [(1+\eta)(c^\dagger_{\bf r+\hat x}+c^\dagger_{\bf r +\hat y}+c^\dagger_{\bf r + \hat z}) \nonumber \\
& &+(1-\eta)(c^\dagger_{\bf r - \hat x}+c^\dagger_{\bf r - \hat y}+c^\dagger_{\bf r - \hat z}) ]c_{\bf r},
\label{eq:3dlattice}
\end{eqnarray}
where $\delta = \pm \hat x \pm \hat y, \pm \hat x \pm \hat z, \pm \hat y \pm \hat z$, $t'=0.1$, and $\eta=0.1$. We also include a magnetic field $\bf B$ along a generic direction (0.48, 0.64,0.6) with amplitude $B=|{\bf B}|=0.02$ ($0.02/2\pi$ magnetic flux quantum per plaquette).

Before analyzing the model with the complex semiclassical theory, we perform an orthogonal transformation to the $(p_1, p_2, p_\parallel)$ coordinates, where $p_\parallel \parallel {\bf B}$ is along the magnetic field. Then, we establish the EOMs following Eq. 1 and $\dot{\bf p}=(\partial \epsilon/\partial {\bf p})\times {\bf B}$ in the main text:
\begin{eqnarray}
\dot{p}_1 &=& B \frac{\partial \epsilon}{\partial p_2} = B (\frac{\partial \epsilon}{\partial p_x}\frac{\partial p_x}{\partial p_2}+\frac{\partial \epsilon}{\partial p_y}\frac{\partial p_y}{\partial p_2}+\frac{\partial \epsilon}{\partial p_z}\frac{\partial p_z}{\partial p_2}),  \label{eq:3DEOM} \\ \nonumber
\dot{p}_2 &=& -B \frac{\partial \epsilon}{\partial p_1} =-B (\frac{\partial \epsilon}{\partial p_x}\frac{\partial p_x}{\partial p_1}+\frac{\partial \epsilon}{\partial p_y}\frac{\partial p_y}{\partial p_1}+\frac{\partial \epsilon}{\partial p_z}\frac{\partial p_z}{\partial p_1}), \nonumber \\
\dot{p}_\parallel &=& 0,
\end{eqnarray}
where $\epsilon({\bf p})$ is the energy dispersion of Eq. \ref{eq:3dlattice}. We are interested in $p_\parallel \in \mathbb{R}$ and $p_1, p_2 \in \mathbb{C}$, whose physics describes a non-divergent plane wave along the $\bf B$ direction and a wave packet in the perpendicular plane.

Finally, for each value of $p_\parallel$ and complex $\epsilon$, we attain closed orbits following the EOMs in Eq. \ref{eq:3DEOM} and the strategy we outlined in the main text and determine whether they satisfy the quantization condition and contribute to the physical spectra. The calculations are relatively straightforward and nowhere costly. The results for the given $\bf B$ and model in Eq. \ref{eq:3dlattice} are summarized in Figs. 2c and 2d in the main text.

\subsection{IIIB. The non-Hermitian skin effect and generalized WKB approximation}

\begin{figure}
\includegraphics[width=0.98\linewidth]{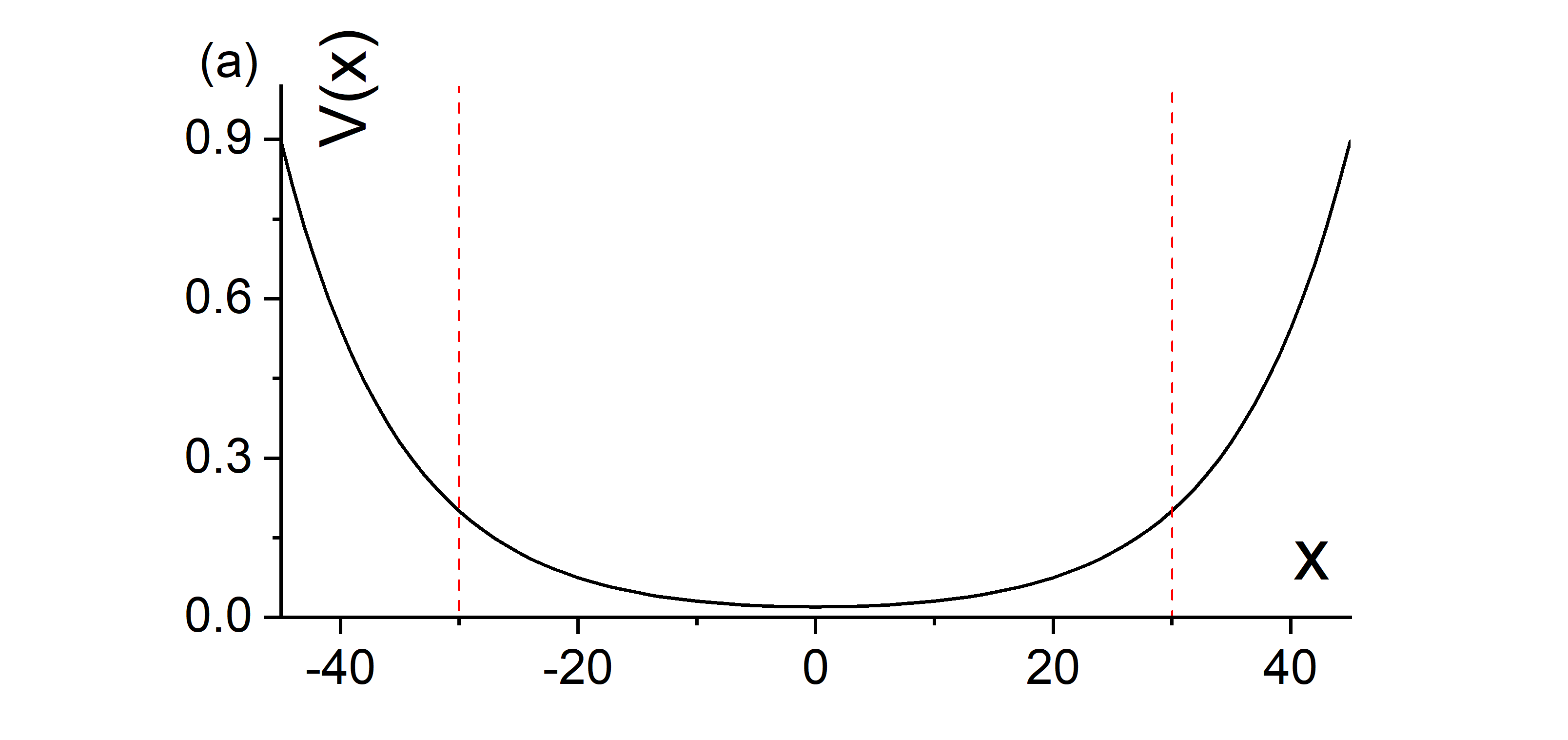}
\includegraphics[width=0.98\linewidth]{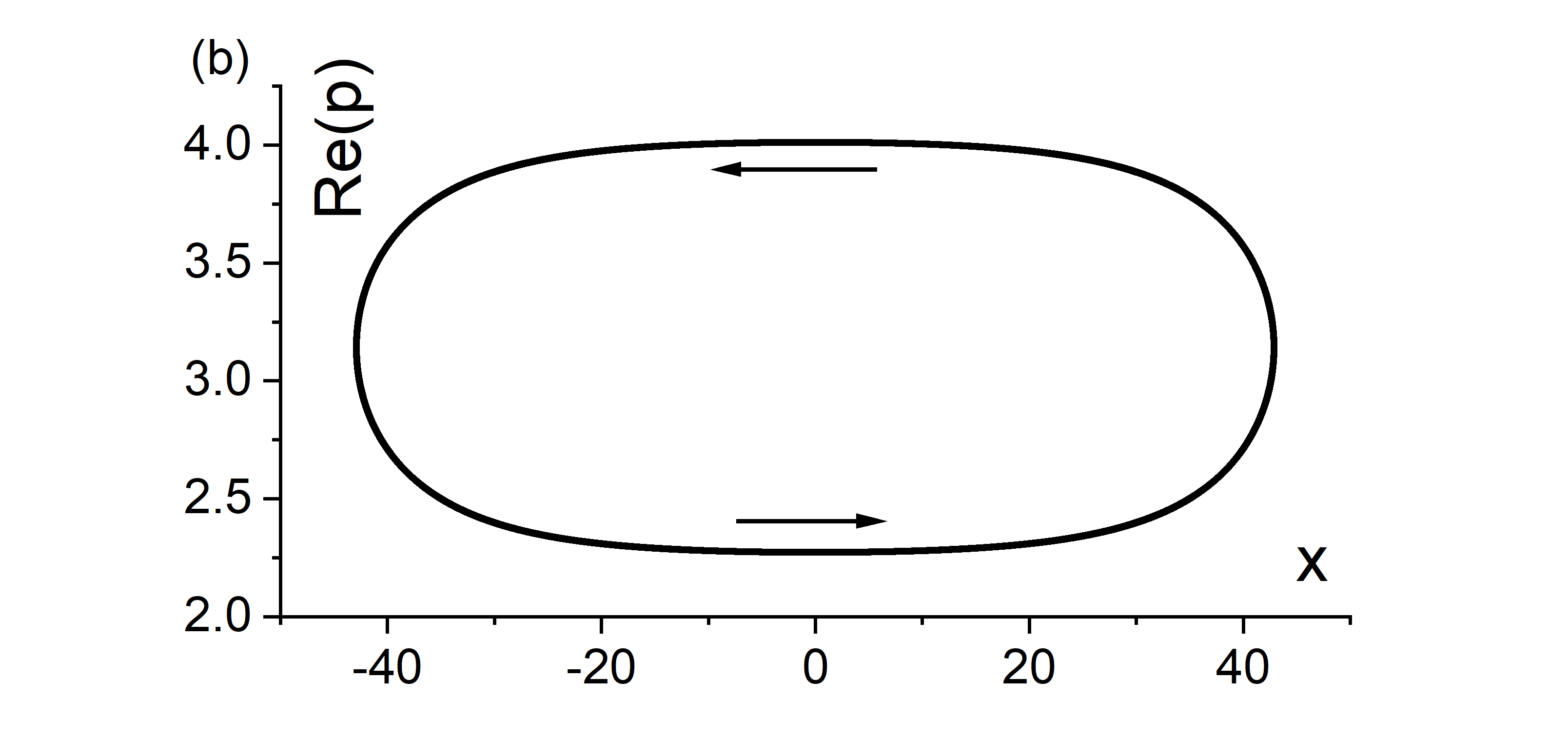}
\includegraphics[width=0.98\linewidth]{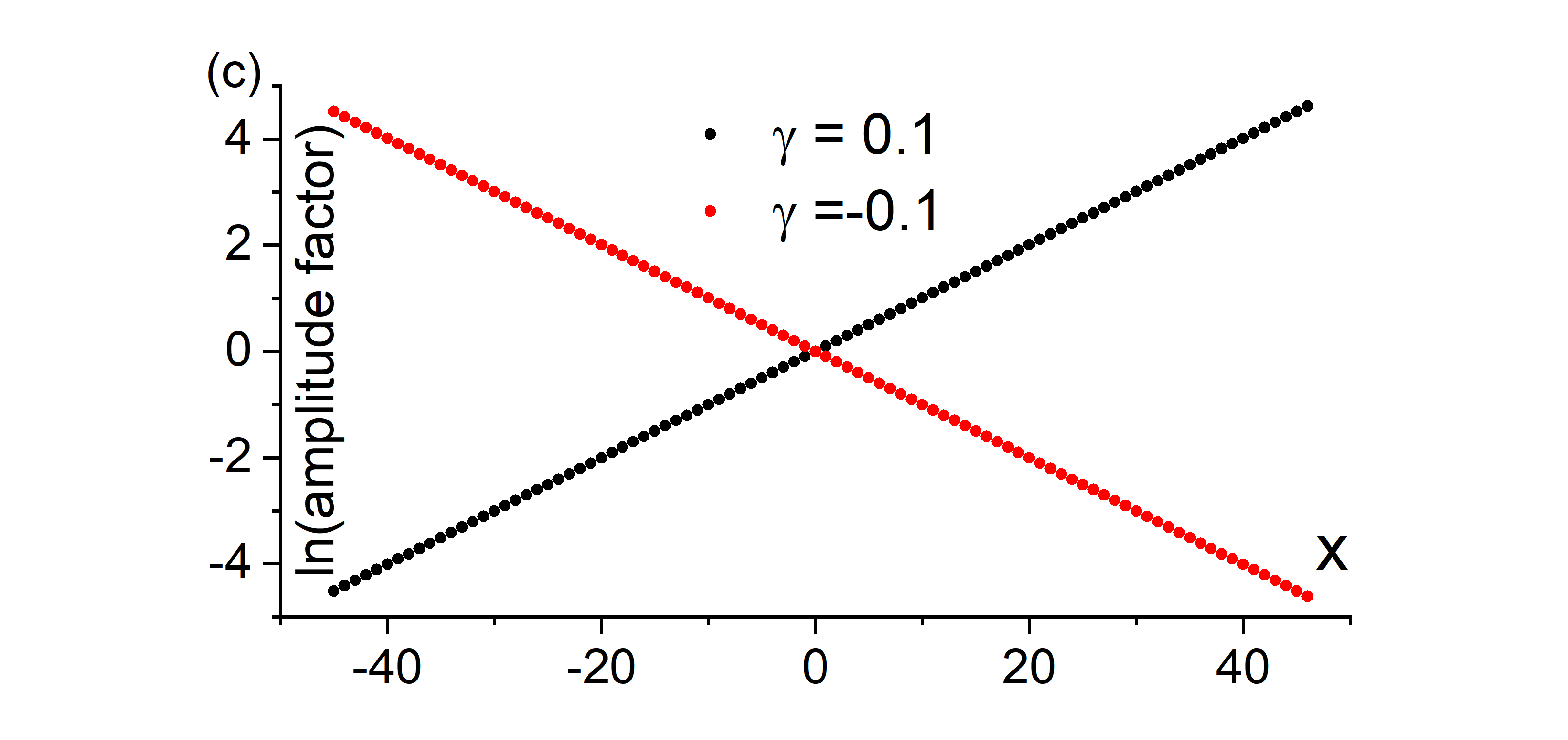}
\caption{(a) The profile of the confining potential in Eq. \ref{eq:appVx} introduces two soft boundaries. The red dashed lines show the setting for $L=60$, beyond which the potential increases more rapidly. (b) The closed orbit following EOMs in Eq. \ref{eq:appEOM_NHSE} retains a real-valued $x$ and a constant ${\rm Im}(p) = \mp 0.1003$ for $\gamma = \pm 0.1$, respectively. The arrows mark the direction of the cyclotron motion. (c) The amplitude-modification factor of the wave packet leads to the emergence of the NHSE. Note the log scale in the vertical direction. }
\label{fig:nhmodel_nhse}
\end{figure}

In this subsection, we apply the complex semiclassical theory to the well-known non-Hermitian Hatano-Nelson model on a 1D lattice:
\begin{eqnarray}
\hat{H} &=& \sum_x (t+\gamma) c^\dagger_{x+1} c_x + \sum_x (t-\gamma) c^\dagger_{x} c_  {x+1} \nonumber\\
&=& \sum_k \left[2t\cos(p) - 2i\gamma\sin(p)\right] c^\dagger_{k} c_k,
\end{eqnarray}
where we set $t=1$ and $\gamma=\pm 0.1$ without loss of generality. In particular, as a soft boundary condition, we introduce the following confining potential:
\begin{equation}
    V(x) = U\left[e^{\alpha(x-L/2)} + e^{-\alpha(x+L/2)}\right],
    \label{eq:appVx}
\end{equation}
whose profile with $U=0.2$, $\alpha=0.1$, and $L=60$ is in Fig. \ref{fig:nhmodel_nhse}a. Note that instead of the common, hard-wall open boundary conditions, we have employed smoother boundaries to suit a semiclassical theory better.

We start with the following energy relation and EOMs for the complex semiclassical theory:
\begin{eqnarray}
    \epsilon &=& 2t\cos(p)+2i\gamma\sin(p) + V(x), \nonumber\\
    \dot x &=& \partial \epsilon / \partial p = -2t\sin(p)-2i\gamma\cos(p), \nonumber \\
    \dot p &=& -\partial \epsilon / \partial x = \alpha U\left[e^{-\alpha(x+L/2)}-e^{\alpha(x-L/2)}\right].
    \label{eq:appEOM_NHSE}
\end{eqnarray}
Interestingly, we can keep $x, t\in \mathbb{R}$ with an initialization that satisfies:
\begin{equation}
    {\rm Im}(x)=0, {\rm Im}(p) = \frac{1}{2} \ln \frac{t-\gamma}{t+\gamma}.
    \label{eq:appInit_NHSE}
\end{equation}
Then, following the EOMs in Eq. \ref{eq:appEOM_NHSE}, $\dot p$ becomes purely real, which keeps ${\rm Im}(p)$ constant with $t \in \mathbb{R}$, and in turn, $\dot x$ remains purely real, Eq. \ref{eq:appInit_NHSE} further holds, and so on so forth. Subsequently, we can obtain a closed orbit; see Fig. \ref{fig:nhmodel_nhse}b for example.

Such a closed orbit with real-valued $x \in \mathbb{R}$ in the complex semiclassical theory is intimately connected with the WKB approximation. In the conventional WKB approximation, the position $x$ remains real, while the momentum $p$ becomes purely real or imaginary depending on $E>V(x)$ or $E<V(x)$, respectively. The semiclassical orbits in Eq. \ref{eq:appInit_NHSE} and Fig. \ref{fig:nhmodel_nhse}b are slightly more general, as they concern a fully complex $p$. In addition, the continuous condition of the (generalized) WKB approximation requires the boundary conditions $\int p\cdot dx = 2\pi n+\gamma$, where $n\in\mathbb{Z}$ and $\gamma = m\pi/2 = \pi$ for $m=2$ turning points along the $\pm x$ directions, consistent with the quantization condition for the complex semiclassical theory in Eq. 2 in the main text.

Notably, the non-Hermitian skin effect (NHSE) emerges naturally from the complex semiclassical theory and the generalized WKB theory. Although the closed orbit ventures across the entire length of the system (Fig. \ref{fig:nhmodel_nhse}b), the contribution of the wave packets to the target wave function receives an extra factor $\exp[-\int {\rm Im}(p) dx]$ contributed by the complex-valued geometrical phase - an exponential weight amplification (attenuation) to the right (left) - a signature of the NHSE (Fig. \ref{fig:nhmodel_nhse}c). The tendency reverses with a sign change of $\gamma$. Similarly, this factor corresponds to $\mathcal{A}$ in Eq. 12 of the main text in the complex semiclassical theory. Indeed, ${\rm Im} (p)$ given by Eq. \ref{eq:appInit_NHSE} contributes a universal factor:
\begin{equation}
\exp (-0.5x \ln \frac{t-\gamma}{t+\gamma}) = (\frac{t+\gamma}{t-\gamma})^{x/2},
\end{equation}
at position $x$, which is consistent with the similarity-transformation embodiment of the NHSE \cite{Yongxu2023bulk}.


\subsection{IIIC. The lattice model with a complex magnetic field $B$}

The complex semiclassical theory allows us to locate interesting quantum physics and phenomena from a novel perspective. For instance, we can re-express the EOMs in Eq. 14 in the main text with three complex variables $p_x$, $p_y$, and $Bt$:
\begin{eqnarray}
\partial p_x / \partial Bt &=& 2V\sin p_y + i\eta e^{i(p_x+p_y)}, \nonumber\\
\partial p_y / \partial Bt &=& -[2 \sin p_x + i\eta e^{i(p_x+p_y)}]. \label{eq:eq16re}
\end{eqnarray}
Since $p_x, p_y, t \in \mathbb{C}$ are complex variables, it is natural to analytically continue $B$ to the complex domain.

\begin{figure}
\includegraphics[width=0.98\linewidth]{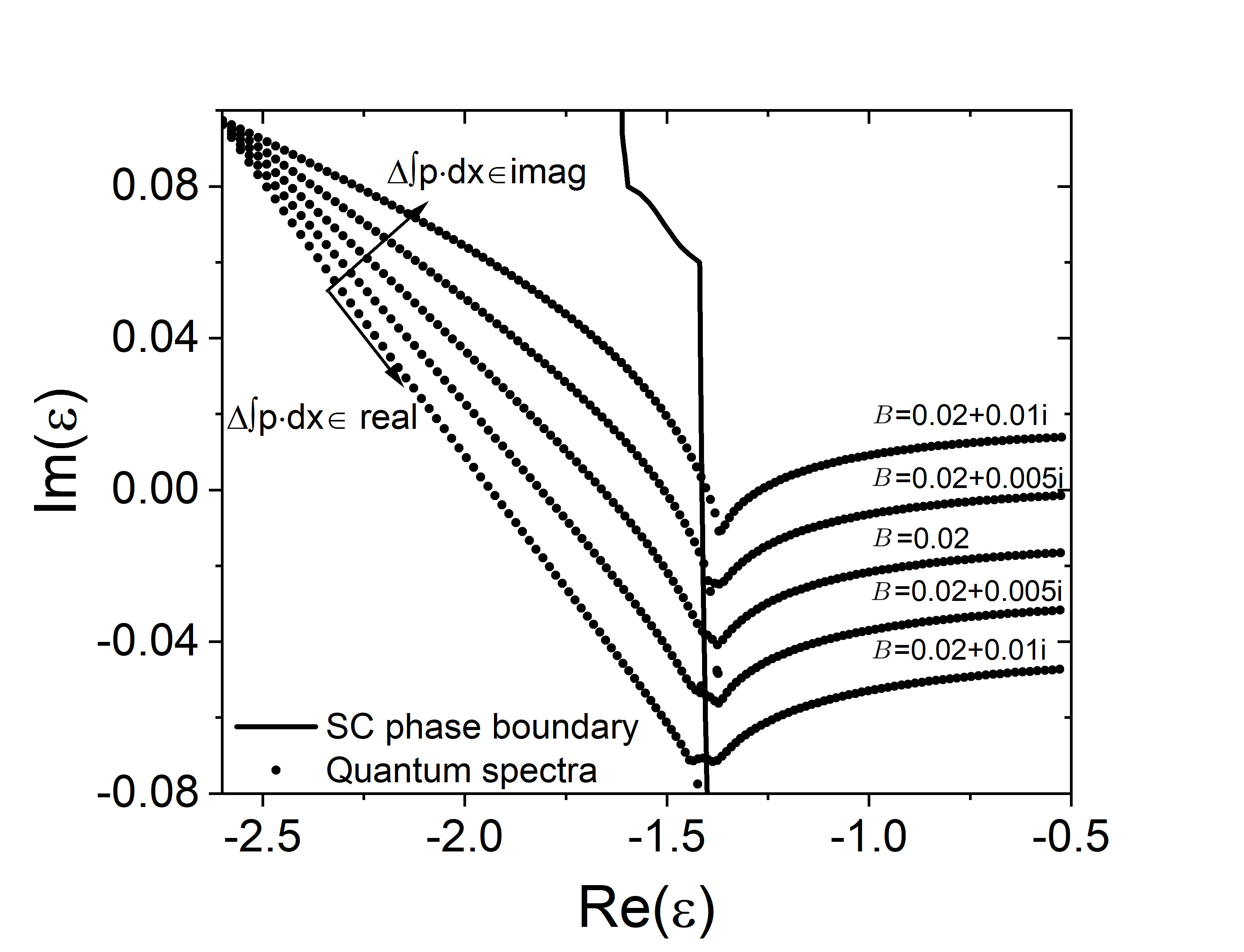}
\caption{The spectra of the non-Hermitian lattice model in Eq. 13 in the main text follow the quantization condition $\oint {\bf p} \cdot d{\bf r} = 2\pi(n+1/2)$. $V=0.3$, $\eta=0.1i$. By varying a magnetic field $B \in \mathbb{R}$, the changes to the geometric phase $\oint {\bf p} \cdot d{\bf r}$ are real-valued, altering the spacing between the energy levels yet keeping the overall envelope curve in the complex $\epsilon$ space. In comparison, a change of $B$ to the complex domain alters the quantization conditions in a nontrivial way that completely shifts the curve and, thus, the whole spectrum. }
\label{fig:lattice_pd2}
\end{figure}

Inserting complex $B$ into the lattice model in Eq. 13 in the main text, we obtain the energy spectra and eigenstates via exact diagonalization. Further, we determine the localization properties and phases via the inverse participation ratios (IPRs), which remain consistent with the complex semiclassical theory, as shown in Fig. 3 in the main text as well as Fig. \ref{fig:lattice_pd2}.

The physics of such a complex magnetic field deserves further study. While a regular, real-valued magnetic field introduces a geometric phase for any path encircling its magnetic flux, a complex magnetic field allows amplification and attenuation - a ``geometric amplitude" effect. Even by itself, this effect is non-unitarity, offering a new origin for non-Hermitian quantum physics, potentially realizable in circuit-model simulators with directional dampers and amplifiers \cite{EzawaPRB2019, EzawaPRB2019b, Helbig2020}.

Such a geometric effect also poses direct and nontrivial consequences on the quantization conditions; see the arrows in Fig. \ref{fig:lattice_pd2}. When we vary a real-valued magnetic field $B$, it alters $\oint {\bf p} \cdot d{\bf r} \in \mathbb{R}$, thus the spacings between and locations of the quantum levels along the original curve in the complex $\epsilon$ space. In comparison, however, a complex $B \in \mathbb{C}$ can alter the imaginary part of $\oint {\bf p} \cdot d{\bf r}$, thus shifting the entire curve elsewhere and completely recasting the spectrum. Such impacts are explicit in the current model example, where $\oint {\bf p} \cdot d{\bf r} = B^{-1} \oint p_x \cdot d(p_y) \propto B^{-1}$.

\section{IV. Wave functions from the complex semiclassical theory}

\subsection{IVA. Geometric phase under complex variables}

Conventionally, the semiclassical theory can provide not only the energy spectrum but also approximate wave functions. Among the contributions over the semiclassical orbit, the geometric phase $\int {\bf p}\cdot d{\bf r}$ plays a vital role \cite{Chaosbook}.

Despite the generalization to the complex domains, the geometric phase in the complex semiclassical theory remains the form of $\oint {\bf p}\cdot d{\bf r}$ along a closed orbit. To see this, we consider a loop around an infinitesimal plaquette $|\delta{\bf p}_c|, |\delta{\bf r}_c|\rightarrow 0$:
\begin{eqnarray}
& &W(0, -\delta{\bf p}_c)W(-\delta{\bf r}_c, 0)W(0, \delta{\bf p}_c)W(\delta{\bf r}_c, 0) \nonumber \\
&=& e^{-i \hat{x}\cdot \delta{\bf p}_c}e^{i\hat{p}\cdot \delta{\bf r}_c}e^{i \hat{x}\cdot \delta{\bf p}_c} e^{-i\hat{p}\cdot \delta{\bf r}_c } = e^{-[\hat{p}, \hat{x}]\delta{\bf p}_c\cdot \delta{\bf r}_c}  \nonumber \\
&=& e^{i\delta{\bf p}_c\cdot \delta{\bf r}_c},
\end{eqnarray}
which follows from the Glauber's theorem. Note we have applied the minus signs instead of the Hermitian conjugates for the inverses, consistent with the biorthogonal convention for generic ${\bf r}_c, {\bf p}_c \in \mathbb{C}$. The action's result is a geometric phase equaling the (complex) volume of the encircled phase space. We can also decompose larger loops from the smaller, more fundamental loops, whose overall geometric phases contribute additively as the total volume $\oint {\bf p}_c\cdot d{\bf r}_c$ of the encircled phase space. We note that the geometric phase from $({\bf r}_c, {\bf p}_c)$ to $({\bf r}_c+\delta {\bf r}_c, {\bf p}_c+ \delta {\bf p}_c)$ is obtainable through the trajectory:
\begin{eqnarray}
   & &\langle {\bf r}_c+\delta {\bf r}_c, {\bf p}_c+ \delta {\bf p}_c|W(\delta{\bf r}_c, \delta{\bf p}_c)|{\bf r}_c, {\bf p}_c\rangle  \\\nonumber &=& \langle 0|W(-{\bf r}_c-\delta {\bf r}_c, -{\bf p}_c-\delta {\bf p}_c) W(\delta{\bf r}_c, \delta{\bf p}_c)W({\bf r}_c, {\bf p}_c)|0\rangle.
\end{eqnarray}

We can obtain the wave functions with such geometric phases augmenting the wave packets, as we demonstrate in the following subsections. As we have defined $|{\bf r}_c, {\bf p}_c\rangle = W({\bf r}_c, {\bf p}_c)|0\rangle$, the states we obtain are the right eigenstates. For the left eigenstates, we need to start from $\langle{\bf r}_c, {\bf p}_c| = \langle 0|W(-{\bf r}_c, -{\bf p}_c)$, instead of simple Hermitian conjugation, for non-Hermitian quantum systems.

In addition, the single-valuedness of the wave function constrains the geometric phase $\oint {\bf p}\cdot d{\bf r}$ along a closed orbit, hence the quantization condition. For scenarios with ${\bf p}\in \mathbb{C}$ and ${\bf r}\in \mathbb{R}$, we have $\gamma=1/2$ by counting the turning points in a generalized WKB approximation (Sec. IIIB), each contributing a $\pi/2$ phase. In the complex semiclassical theory over more generic scenarios ${\bf r}, {\bf p}\in \mathbb{C}$, the definition of turning points may become obscure, yet we may still locate the value for $\gamma$ following the duality transformation ${\bf r}\rightarrow {\bf p}$ and ${\bf p}\rightarrow -{\bf r}$, which contributes a $\pi/4$ phase each and totals $\pi$ around a closed orbit \cite{Chaosbook}.

\subsection{IVB. Analytical derivations of eigenstates of continuous models}

In this subsection, we demonstrate the complex semiclassical theory on the derivation of the quantum eigenstates for the non-Hermitian continuous model in Eq. 3 in the main text. Such a model is exactly solvable with the ladder operators in Eq. 4 in the main text, and without loss of generality, we set $2\alpha=\omega-i\eta$, $2\beta=\omega+i\eta$, $\kappa = \eta/i\omega$ for simpler expressions. Subsequently, $\hat a \propto \hat a_0$, $\hat b^\dagger \propto \hat a_0^\dagger + \kappa \hat a_0$, and the right eigenstates $\hat a|0\rangle = 0$, and $|n+1\rangle \propto \hat{b}^\dagger|n\rangle$ attain the expressions:
\begin{equation}
|0\rangle = |0\rangle_0, \; |n\rangle \propto (a_0^\dagger + \kappa \hat a_0)^n |0\rangle_0, \label{eq:wfbench}
\end{equation}
which serve as our benchmarks. Here, $\hat a_0 = (\hat x+i\hat p)/\sqrt{2}$ and $a^\dagger_0 a_0|m\rangle_0 = m|m\rangle_0$ are based upon a Hermitian, isotropic harmonic oscillator.

Now, we turn to the complex semiclassical theory, whose closed orbits for the non-Hermitian quantum model are Eq. 5 in the main text. As we show in the Sec. VA, the complex variables have redundancy in their conventions, and orbits that differ by a mapping $z'_1=z_1 \cdot \Delta z$, $z'_2 =z_2 / \Delta z$ have the same physical properties and are equivalent to each other. We thus focus on the limit $z_1 \rightarrow 0$, where the subsequent orbit $x(t)=z_2 e^{-i\omega t}$, $p(t)=z_2 e^{-i\omega t} \alpha/i\beta$ becomes large $|z_2| \rightarrow \infty$ and highly isotropic. The symmetry allows us to distribute the geometric phase along the orbit evenly and regard the shape of the wave packet as isotropic: $|0\rangle_0$. Further, the wave packet's COM location (momentum) and $\mathcal{A}$ factor evolve as:
\begin{eqnarray}
{x_r(t)+ip_r(t)}&=& (1+\frac{\alpha}{\beta}) z_2 e^{-i\omega t}, \nonumber\\
\mathcal{A}(t)&\propto& \exp [(\frac{\alpha^2}{\beta^2}-1)z^2_2 e^{-i2\omega t}],
\end{eqnarray}
following Eqs. 11 and 12 in the main text \footnote{Note $x^2_r(t)+p_r^2(t)=|z_2(1+\alpha/\beta)|$ is constant.}.

Summing over the wave packet $\mathcal{A}(t)|x_r(t), p_r(t)\rangle$ with the geometric phase $\exp(in\omega t)$ over the closed orbits, we obtain:
\begin{eqnarray}
|n\rangle &\propto& \int_0^T dt \cdot \exp(in\omega t) \cdot \mathcal{A}(t)W(x_r(t), p_r(t))|0\rangle_0 \nonumber \\
&\propto& \oint dz \cdot z^{-n-1} \cdot \exp[ \frac{z^2}{4} (\frac{\alpha^2}{\beta^2}-1)+\frac{z}{\sqrt{2}}(1+\frac{\alpha}{\beta}) \hat a^\dagger_0]|0\rangle_0 \nonumber \\
&\propto& \oint dz \cdot z^{-n-1} \cdot e^{\kappa z^2/2+z \hat a^\dagger_0}|0\rangle_0,
\label{eq:qstate_nH}
\end{eqnarray}
where $z = z_2 e^{-i\omega t}$. We have renormalized $(1+\alpha/\beta)z/\sqrt{2} \rightarrow z$ in the third line. Note $\kappa = (\alpha-\beta)/(\alpha+\beta) = \eta/i\omega $ in our setting. Following the residual theorem, the contour integral returns the $z^n$'s coefficient in the series expansion of $\exp (x\hat{a}^\dagger + \kappa x^2/2)$, which compares consistently with the benchmarks in Eq. \ref{eq:wfbench} as Table I in the main text.

\subsection{IVC. Numerical derivation of wave functions of lattice models}

More generally, however, we cannot attain the wave functions analytically as in the last subsection, e.g., due to the absence of a solvable orbit expression or subsequent integration. Consequently, we need to resort to numerical calculations. In this subsection, we derive the wave functions, i.e., $\psi_n(x)$ in the real space, from the complex semiclassical theory for the non-Hermitian lattice model in Eq. 13 in the main text. By choosing the Landau gauge $A_x=0, A_y=Bx$, the Hamiltonian in $\hat x$ (for each decoupled $k_y$) takes the form:
\begin{eqnarray}
\hat{H}&=&\sum_{x} (\eta e^{iBx}-1) c^\dagger_{x+1}c_{x} -c^\dagger_{x} c_{x+1} -2V \cos(Bx) c^\dagger_{x} c_{x}, \nonumber \\ \label{eq:appstateham}
\end{eqnarray}
which we diagonalize for its energy eigenvalues and eigenstates. We set $B = 0.005$, $V = 0.3$. Without loss of generality, we choose its $n=3$ eigenstate as a benchmark, which is sufficiently localized, and the non-Hermitian error accumulation does not pose a severe issue.

\begin{figure}
\includegraphics[width=0.98\linewidth]{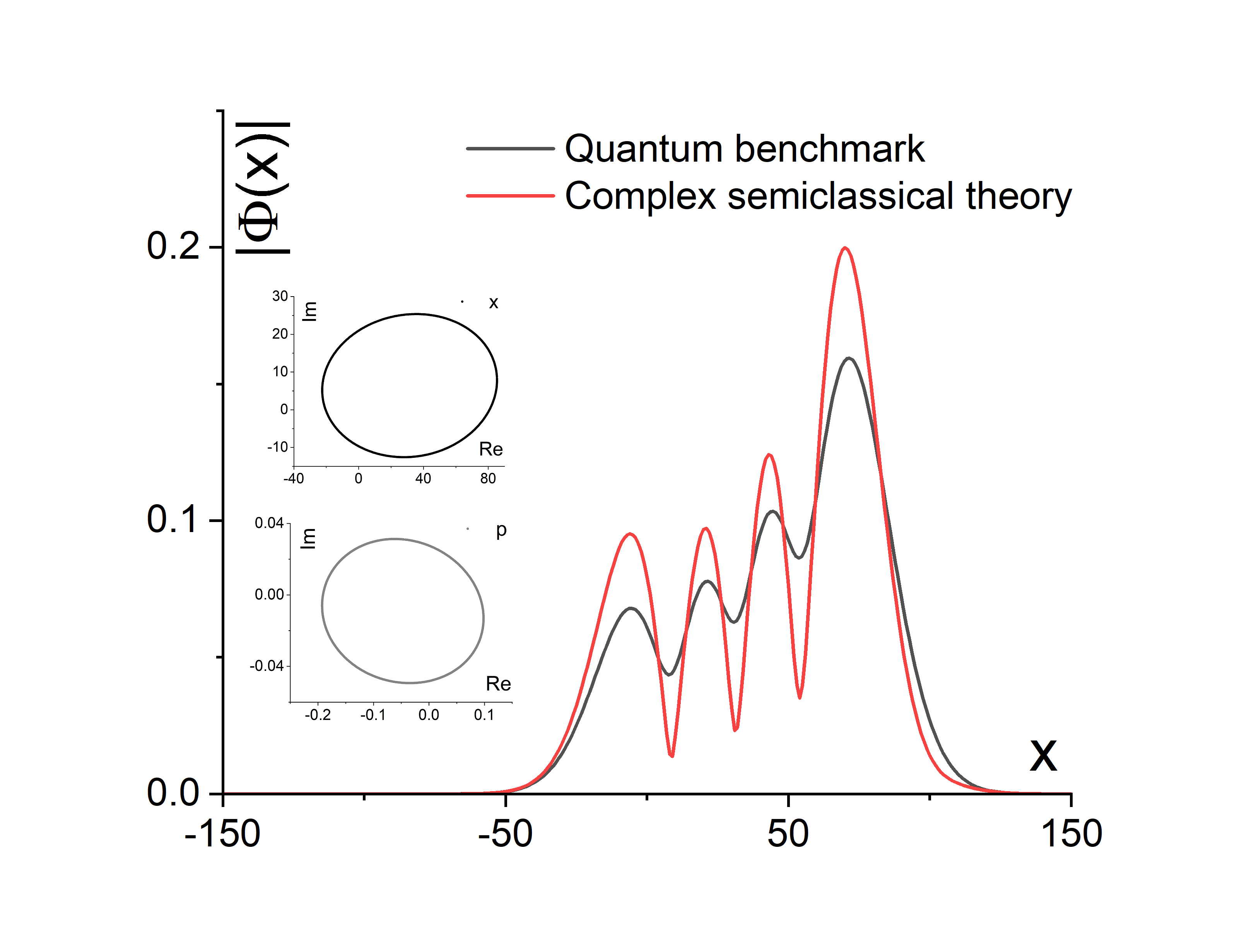}
\caption{The wave function $\psi_3(x)$ from the complex semiclassical theory for the non-Hermitian model in Eq. \ref{eq:appstateham} and the corresponding quantum eigenstate compare qualitatively well. $B = 0.005$ and $V = 0.3$. The energy eigenvalue $\epsilon_3= -2.5905+0.0959i$ is consistent between the complex semiclassical theory and the quantum benchmark. The insets show the orbit in the complex $x$ and $p$ spaces. }
\label{fig:appstate}
\end{figure}

From the complex semiclassical theory, on the other hand, we begin with a closed orbit, $x_c(t)$ and $p_c(t)$, with the quantization condition $\oint p_c \cdot dx_c = 7\pi = 2\pi(n+1/2)$. We illustrate the orbit in the complex $x$ and $p$ spaces in the inset of Fig. \ref{fig:appstate}. Its energy $\epsilon_3= -2.5905+0.0959i$ is consistent with the quantum benchmark.

Similar to Eq. \ref{eq:qstate_nH}, we obtain the target wave function as:
\begin{eqnarray}
    \psi_3(x)&\propto& \int_0^T dt \cdot \exp(6\pi it/T)\cdot \mathcal{A}(t) \cdot \phi_{x_c(t)+ip_c(t)}(x),
\end{eqnarray}
where $\mathcal{A}(t)$ and $x_c(t)+ip_c(t)=x_r(t)+ip_r(t)$ are the multiplying factor and COM position $x_r(t)$ (momentum $p_r(t)$). We obtain $\phi_{x_c(t)+ip_c(t)}(x)$ approximately by translating the ground state wave-packet $\phi_0(x)$. After numerical integration, the real-space wave function $\psi_3(x)$ compares qualitatively consistently with the quantum benchmark (Fig. \ref{fig:appstate}).

We note that given the orbits in the complex semiclassical theory, the analytical and numerical approaches for quantum eigenstates and wave functions are straightforwardly generalizable to other non-Hermitian quantum systems.

\section{V. Complex-variable and biorthogonal conventions in the complex semiclassical theory}

\subsection{VA. Complex-variable convention and redundancy}

In Eqs. 11 and 12 of the main text, we have established the connection between the complex variables $x_c, p_c \in \mathbb{C}$ and the wave packet $\mathcal{A}|x_r, p_r\rangle$ with real variables $x_r, p_r \in \mathbb{R}$ and an additional factor $\mathcal{A} \in \mathbb{C}$. Therefore, even for the same $x_r, p_r \in \mathbb{R}$, there exists a redundancy in $x_c, p_c$ that follows the equality $x_r+ip_r = x_c+ip_c$ and differs by their $\mathcal{A}$ factors, which can be attributed to quantum states' normalization and phase conventions. In this subsection, we demonstrate that the multiple orbits under different conventions $\mathcal{A}\rightarrow \mathcal{A}\cdot \Delta\mathcal{A}$ describe physics consistently.

\begin{figure}
\includegraphics[width=0.49\linewidth]{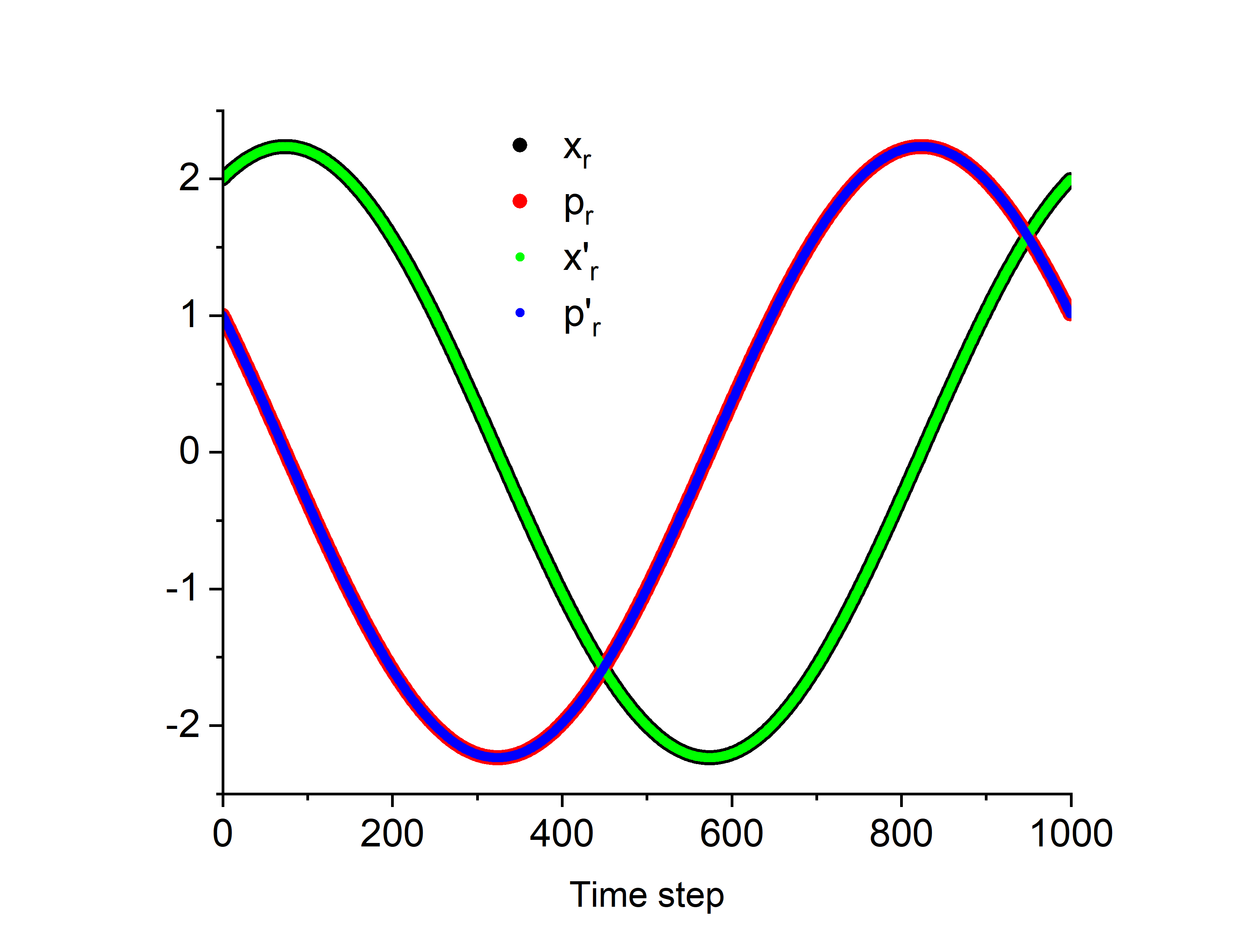}
\includegraphics[width=0.49\linewidth]{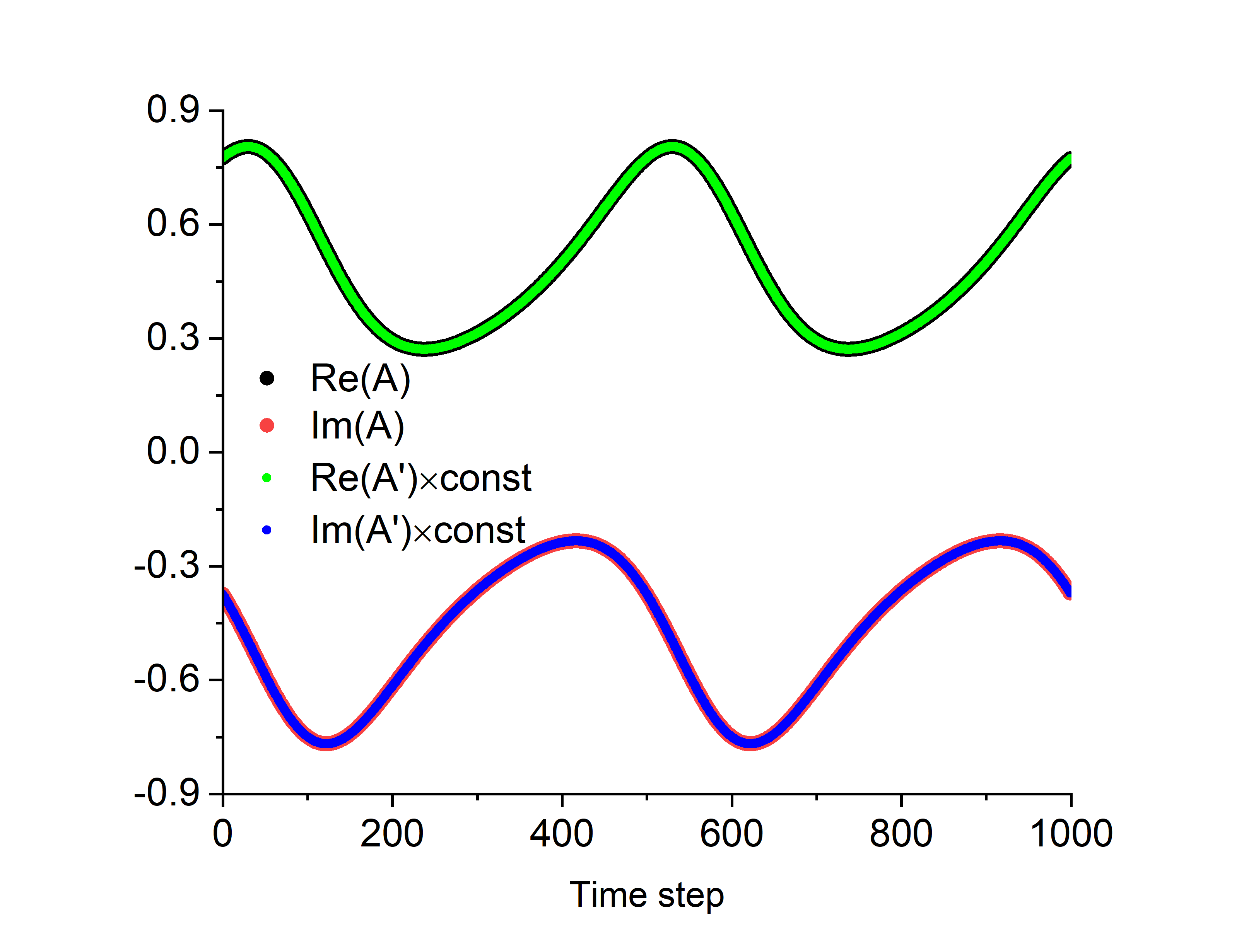}
\caption{We initialize two semiclassical orbits with $x=2, p=1$ and $x'=2.3+0.3i, p'=0.7+0.3i$, which correspond to the same $x_r, p_r \in \mathbb{R}$, and (left) their wave packets' COMs evolve consistently, while (right) the ratio $\Delta\mathcal{A}$ between their $\mathcal{A}$ factors remains constant, therefore constitute merely a difference in their conventions. }
\label{fig:convcons}
\end{figure}

For example, we consider the continuous model in Eq. 3 in the main text and map out the trajectories following various initializations, which correspond to the same $x_r, p_r \in \mathbb{R}$ with identical energy $\epsilon \in \mathbb{C}$. Their subsequent evolution is universal and consistent with each other, see Fig. \ref{fig:convcons}, differing merely in their conventions - the ratios $\Delta \mathcal{A}$ between their $\mathcal{A}$ factors remain constant through their entire orbits. The results and conclusions generalize to other models and $\mathcal{A}$ conventions straightforwardly. Thus, the redundancy over complex variables does not lead to self-contradiction in the complex semiclassical theory.

\subsection{VB. Biorthogonal convention and expectation value in non-Hermitian quantum systems }

We can diagonalize a non-Hermitian Hamiltonian as $\hat{H}=\sum_{n}\epsilon_{n}\ket{nR}\bra{nL}$, where $\ket{nR}$ ($\bra{nL}$) is the right (left) eigenstate with respect to the energy eigenvalue $\epsilon_{n}$. Correspondingly, the time evolution operator is:
\begin{equation}
U\left(t\right)=\exp{(-i\hat{H}t)}=\sum_{n}\exp{\left(-i\epsilon_{n} t\right)}\ket{nR}\bra{nL}.
\end{equation}
Alternatively, we can write the diagonalized $\hat{H}$ in a more compact matrix form as $\Lambda=V^{-1}\hat HV$, where the $n^{th}$ column (row) of $V$ ($V^{-1}$) is exactly $\ket{nR}$ ($\bra{nL}$). $\epsilon_{n}$, on the diagonal $\Lambda$, is the biorthogonal expectation value of $\hat H$: $\epsilon_{n}=\braket{nL|\hat{H}|nR}$. Consistently, it is natural to deduce that the expectation value of an operator $\hat{O}$ is the biorthogonal counterpart, $\braket{\hat{O}}_n=\braket{nL|\hat{O}|nR}$, or more generally, $\braket{\hat{O}}_{nm}=\braket{nL|\hat{O}|mR}$.

Given an initial quantum state $\ket{\Phi(0)}$, we can decompose it in the biorthogonal basis $\ket{\Phi(0)}=\sum_{n}\ket{nR}\braket{nL|\Phi(0)}$. It evolves in time as:
\begin{align}
    \ket{\Phi\left(t\right)}&=U\left(t\right)\ket{\Phi(0)}\nonumber\\
    &=\sum_{nm}\exp{\left(-iE_nt\right)}\ket{nR}\braket{nL|mR}\braket{mL|\Phi(0)}\nonumber\\
    &=\sum_{n}\exp{\left(-iE_nt\right)}\ket{nR}\braket{nL|\Phi(0)}, \label{eq:rstateevolve}
\end{align}
where we have used the property $\braket{nL|mR}=\delta_{nm}$ of the biorthogonal basis. Similarly, we have its bra $\langle \Phi(0)|$'s time evolution as:
\begin{align}
\bra{\Phi(t)}&=\bra{\Phi(0)}U^{-1}(t)\nonumber\\
&=\sum_{n}\exp{\left(iE_nt\right)}\braket{\Phi(0)|nR}\bra{nL},
\end{align}
whereas the Hermitian conjugation of $\ket{\Phi(t)}$ is different:
\begin{align}
\bra{\Phi'(t)}=\sum_{n}\exp{\left(iE_{n}^{*}t\right)}\braket{\Phi(0)|nL}\bra{nR},
\end{align}
as directly follows from Eq. \ref{eq:rstateevolve}.

Thus, the amplitude of $\braket{\Phi(t)|\Phi(t)}$ is conserved, while that of $\braket{\Phi'(t)|\Phi(t)}$ grows or decays exponentially with respect to $t$:
\begin{align}
    \braket{\Phi(t)|\Phi(t)}&=\braket{\Phi(0)|U^{-1}(t)U(t)|\Phi(0)}=1,\nonumber\\
    \braket{\Phi'(t)|\Phi(t)}&=\braket{\Phi(0)|U^{\dagger}(t)U(t)|\Phi(0)}\nonumber\\
    &=\sum_{mn}\exp{\left[i(E_{m}^{*}-E_{n})t\right]}\braket{\Phi(0)|mL} \nonumber\\
    &\qquad\quad\times\braket{mR|nR}\braket{nL|\Phi(0)}.
\end{align}
Therefore, the former convention is more convenient for analyzing inner products and expectation values in non-Hermitian quantum systems.

Employing the biorthogonal convention, we define and evaluate the time evolution of observable $\hat{O}$'s expectation value as:
\begin{align}
\braket{\hat{O}(t)}=\braket{\Phi(t)|\hat{O}|\Phi(t)}=\braket{\Phi(0)|U^{-1}(t)\hat{O}U(t)|\Phi(0)},
\end{align}
which converts back to the common definition in Hermitian systems:
\begin{align}
\braket{\hat{O}(t)}&=\braket{\Phi(0)|U^{-1}(t)\hat{O}U(t)|\Phi(0)}\\\nonumber
&=\braket{\Phi(0)|U^{\dagger}(t)\hat{O}U(t)|\Phi(0)}=\braket{\Phi(t)|\hat{O}|\Phi(t)}.
\end{align}

Whether the expectation value in non-Hermitian quantum systems should be defined and calculated within the biorthogonal basis or merely the right eigenstates remains a subtle matter. When we speak of the NHSE, the focus is solely on (the amplitude of) the right eigenstates: $\braket{\psi'(x)|\psi(x)}$; in contrast, however, the NHSE disappears when we employ the biorthogonal basis: $\braket{\psi(x)|\psi(x)}$. Even when it comes to many-body non-Hermitian quantum systems, these two views are both present in the literature: the spatial density of an $n$-fermion state $\ket{\Psi_{\mu}}$ is calculated via $\braket{\Psi'_{\mu}|\hat{n}_{x}|\Psi_{\mu}}$ with the right eigenstates in Ref. \cite{PhysRevB.102.081115}, while the non-Hermitian many-body bulk polarization $P^{LR}$ takes the biorthogonal form \cite{PhysRevB.101.121109}.

\section{VI. Complex semiclassical theory for dissipative system with friction}

In this subsection, we apply the complex semiclassical theory to a damped harmonic oscillator in 1D with $V(x)=x^2$:
\begin{eqnarray}
    \dot{x} &=& p, \nonumber\\
    \dot{p} &=& -x-\mu p, \label{eq:eomsfric}
\end{eqnarray}
where the coefficient $\mu$ characterizes a friction $F_\mu =-\mu p$. The quantum theory of such a dissipative system has been in continual investigation and controversy for decades and commonly requires external reservoirs to handle the dissipation properly \cite{Philbin_2012}.

The EOMs in Eq. \ref{eq:eomsfric} are linear differential equations and exactly solvable:
\begin{eqnarray}
    x(t) &=& z_1 e^{-c_1 t} + z_2 e^{-c_2 t}, \nonumber\\
    p(t) &=& -z_1 c_1 e^{-c_1 t} - z_2 c_2 e^{-c_2 t}, \label{eq:fricorbit}
\end{eqnarray}
where $c_{1/2}=(\mu \pm \sqrt{\mu^2 -4})/2$ and $z_1$ and $z_2$ are constants determined by the initial conditions. When $\mu < 2$, $c_{1/2}$ are partially imaginary, and the system remains underdamped and residual oscillatory; when $\mu > 2$, $c_{1/2}$ is fully real, and the system becomes overdamped.

\begin{figure}
\includegraphics[width=0.49\linewidth]{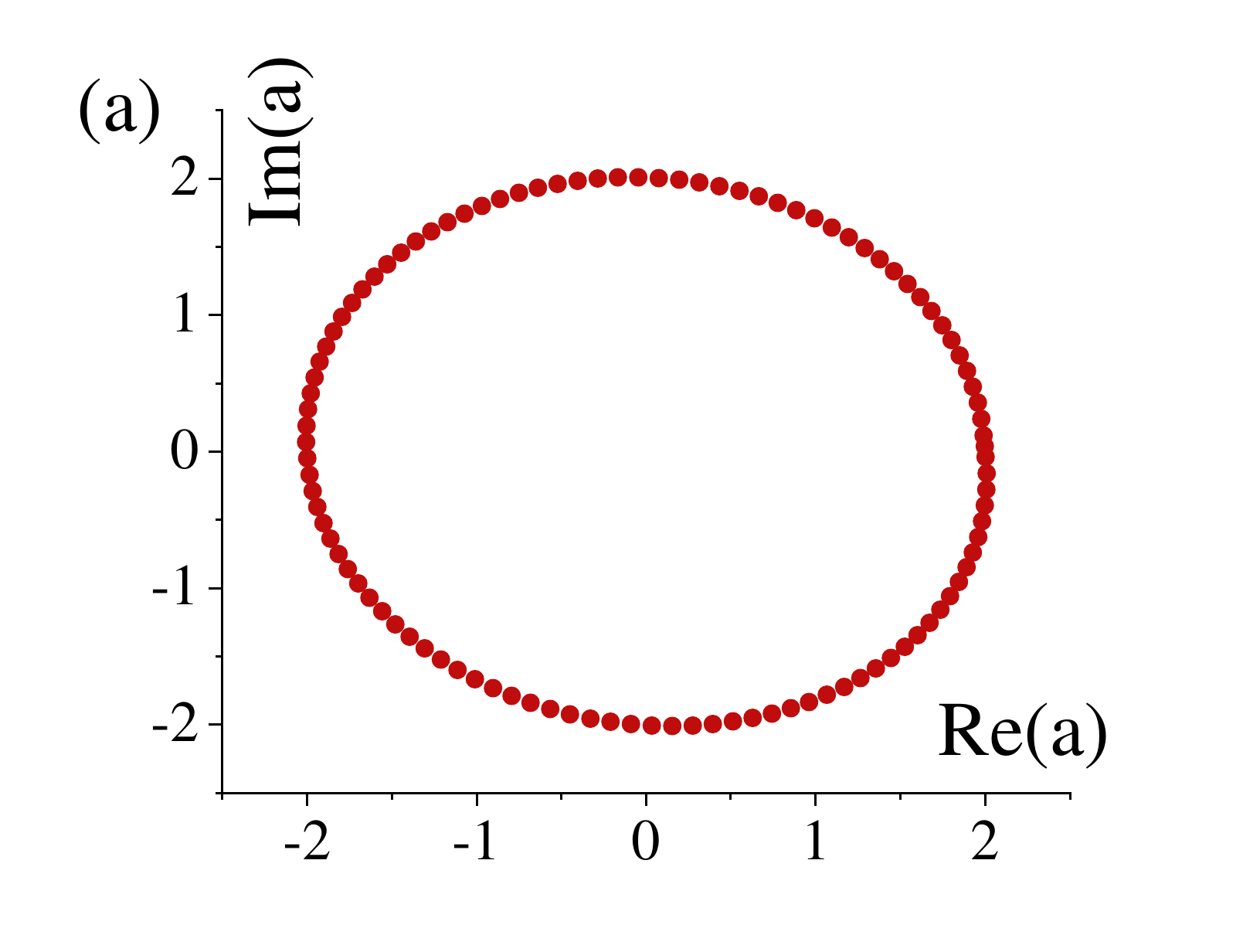}
\includegraphics[width=0.49\linewidth]{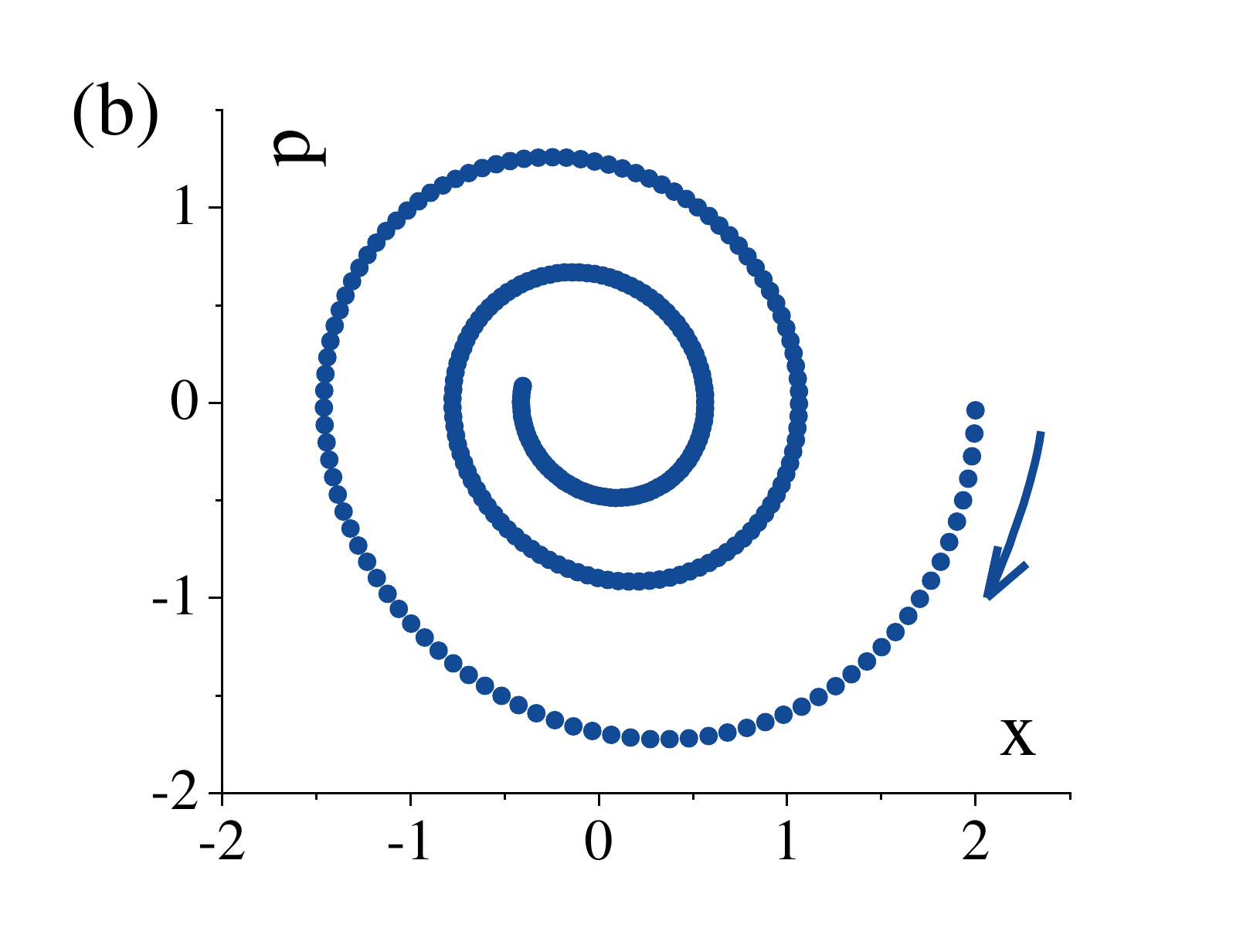}
\includegraphics[width=0.49\linewidth]{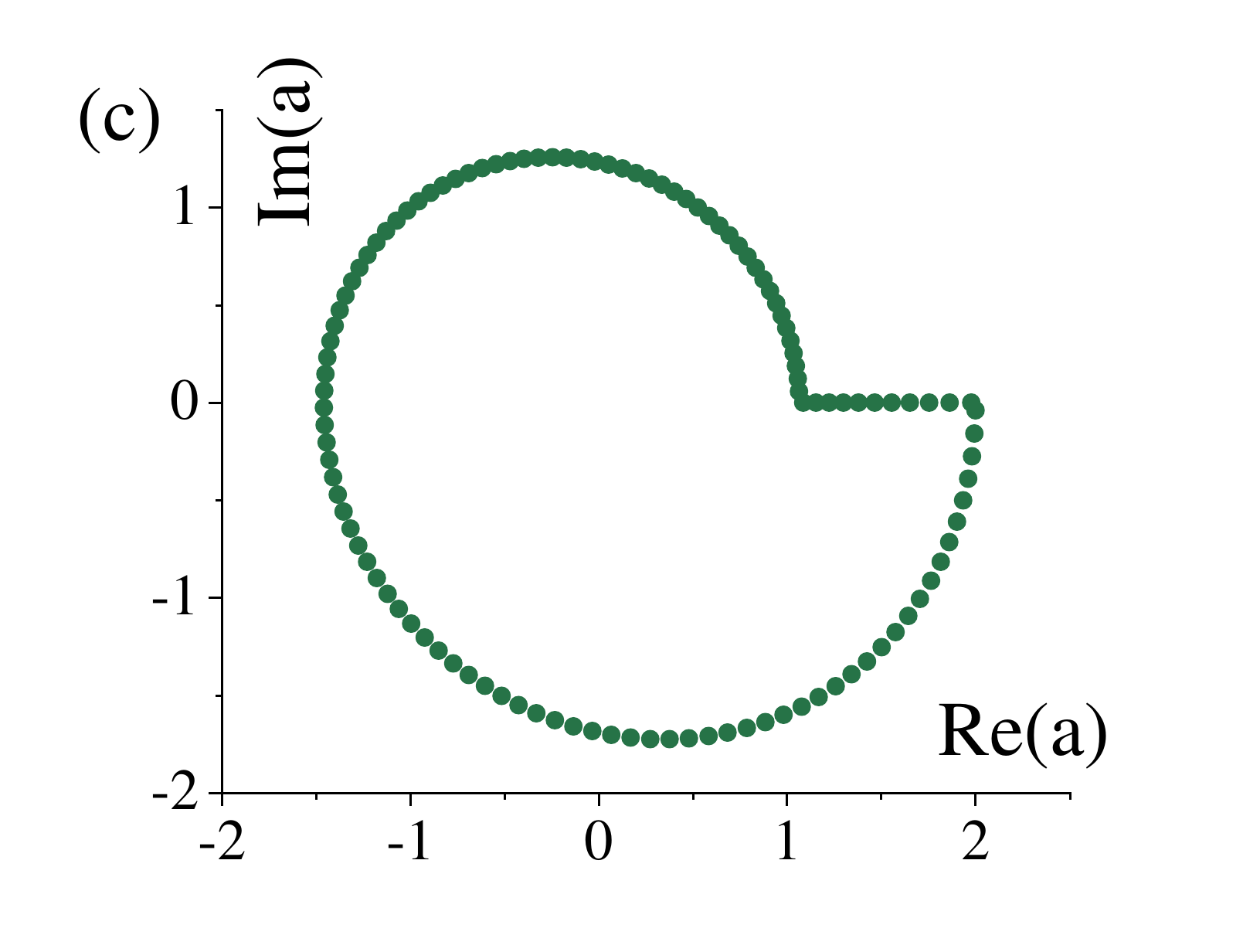}
\includegraphics[width=0.49\linewidth]{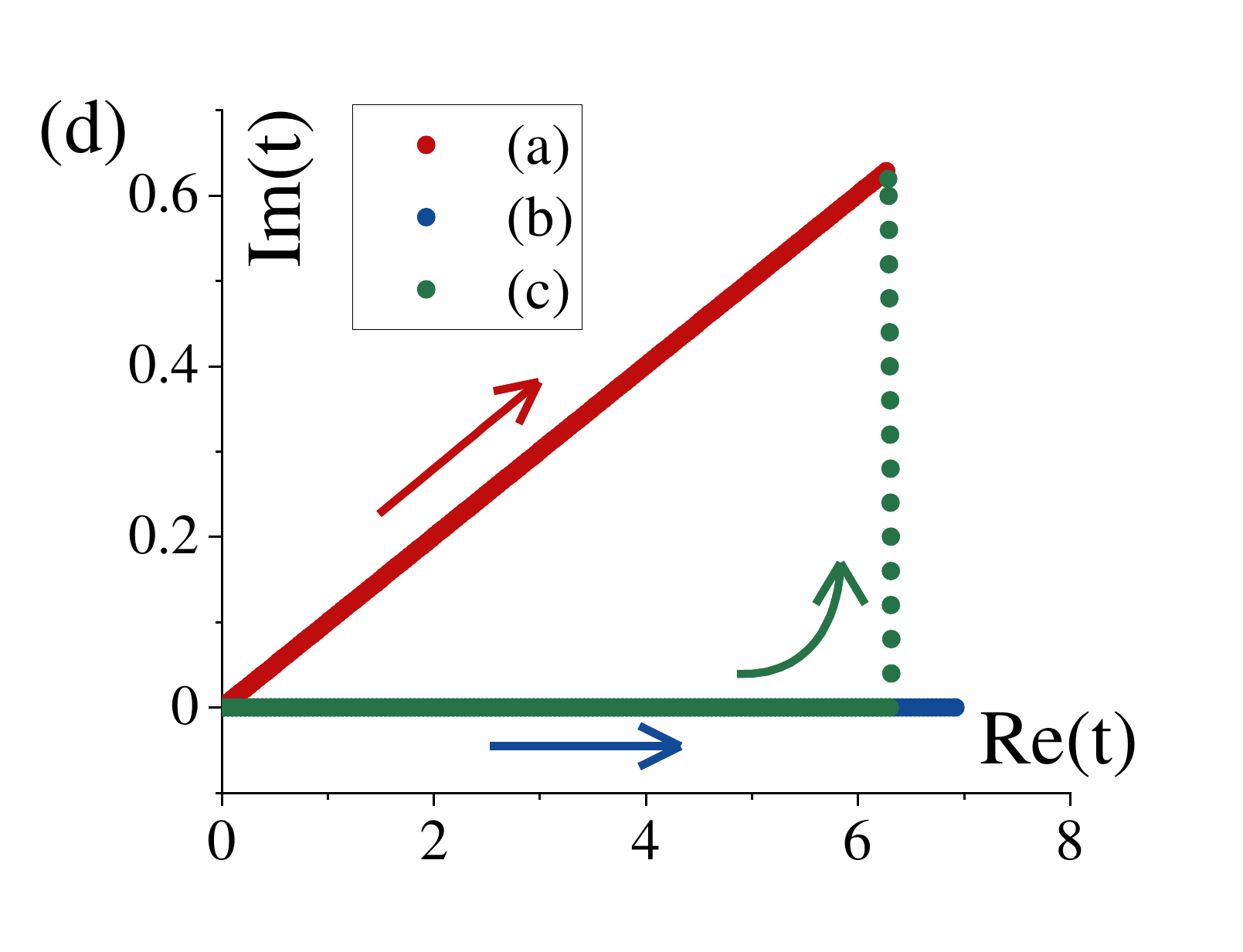}
\caption{Following the EOMs, the semiclassical trajectories of the model in Eq. \ref{eq:eomsfric} with a friction coefficient $\mu=0.2$ form closed loops (a) along the $t/T \in \mathbb{R}$ direction and end up in open spirals (b) along the $t \in \mathbb{R}$ direction in the complex $a$ plane, $a=x+ip$. Without knowing $T$, we may still attain $a(T)=a(0)$ by strategically adjusting $dt$, and the resulting trajectory is in (c). (d) The corresponding trajectories in the complex $t$ plane show the equivalence between different closed orbits reaching the same complex period $T$. $\mu=0.2$ and $T=6.28+0.63i \approx 2\pi i/[(\mu+\sqrt{\mu^2-4})/2]$. }
\label{fig:friction}
\end{figure}

Setting $\mu < 2$ and $z_1 = z_2 = x_0/2$, we obtain the familiar classical behavior of a damped oscillator:
\begin{eqnarray}
    x(t) &=& x_0 e^{-\mu t/2} \cos(\omega t), \nonumber \\
    p(t) &=& - x_0 e^{-\mu t/2} \left[\omega \sin(\omega t)+ \mu \cos(\omega t)/2  \right],
\end{eqnarray}
where $\omega = \sqrt{1-\mu^2/4}$. Due to the extra damping factor $e^{-\mu t/2}$, the absence of closed orbit in real $t\in \mathbb{R}$ is apparent; see Fig. \ref{fig:friction}b.

In the complex semiclassical theory, on the other hand, the variables $z_1$, $z_2$, $x$, $p$, and, importantly, $t$ may take on complex values: $t \in \mathbb{C}$. Consequently, we can establish closed orbits in the complex time domain, e.g., $t \parallel T$ and $ic_1 t \in \mathbb{R}$; see Fig. \ref{fig:friction}a. More generally, we can obtain closed orbits numerically via finite-time steps following the EOMs in Eq. \ref{eq:eomsfric} and the strategy we propose in the main text; see Fig. \ref{fig:friction}c. As long as the winding is identical, closed orbits are equivalent, e.g., with the same period $T \approx 2\pi i/c_1$ (Fig.  \ref{fig:friction}d).

Unlike the aforementioned dissipation-less models in the main text and the supplemental materials, the EOMs in Eq. \ref{eq:eomsfric} introduce energy losses and, in general, cannot return both $x$ and $p$ to their initial values. Instead, we demand $a(T) = a(0)$, $a = x + ip = x_r + ip_r$ return to its initial value, so that the real-valued COM position and momentum of the wave packet $x_r, p_r \in \mathbb{R}$ return to their initial values. Nevertheless, the additional factor $\mathcal A$ augmenting the wave packet, as in Eq. 12 in the main text, differs in general. Accordingly, we need to modify the quantization condition so that the geometric phase $\oint {\bf p}\cdot d{\bf r}$ upon cycling around the orbit compensates for the difference. Further studies are fascinating yet beyond the scope of the current work, and we leave them for future research.

\section{VII. Complex semiclassical theory with the Berry curvature}

In the main text, we have established the semiclassical equation of motion (EOM) for the complex variable ${\bf r}_c$ in the absence of the Berry curvature $\bf \Omega$, whose effects we analyze here. Similar to the main text, we begin with the evolution of the wave packet $|{\bf r}_c(t+\tau), {\bf p}_c(t+\tau)\rangle = e^{-i\hat{H}\tau}|{\bf r}_c(t), {\bf p}_c(t)\rangle$ after a short time step $\tau \rightarrow 0$:
\begin{eqnarray}
\dot{\bf r}_c(t) &=& \left[{\bf r}_c(t+\tau) - {\bf r}_c(t)\right]/\tau \label{eq:TR_deri} \nonumber\\
&=& \langle {\bf r}_c(t), {\bf p}_c(t)| e^{i\hat H \tau} \hat{\bf r} e^{-i\hat{H}\tau} - \hat{\bf r}|{\bf r}_c(t), {\bf p}_c(t) \rangle / \tau \\
&=& -i\langle 0| \hat W(-{\bf r}_c(t), -{\bf p}_c(t)) \left[\hat{\bf r}, \hat{H}\right] \hat W({\bf r}_c(t), {\bf p}_c(t))|0\rangle \nonumber \\
&=& \langle 0|\left(\frac{\partial \hat H}{\partial \hat{\bf p}}+\frac{\partial \hat H}{\partial \hat{\bf r}}\times {\bf\Omega}\right)(\hat{\bf r}+{\bf r}_c(t), \hat{\bf p}+{\bf p}_c(t))|0\rangle, \nonumber
\end{eqnarray}
where for the last line, we note that in addition to $[r_i, p_j]=i\delta_{ij}$, we have $[r_i, r_j]=i\epsilon_{ijk}\Omega_k$ with the Berry curvature $\Omega_k=i\epsilon_{ijk}\left(\partial_{k_i}\langle\psi|\partial_{k_j}|\psi\rangle-\partial_{k_j}\langle\psi|\partial_{k_i}|\psi\rangle\right)$. For reasons we have stated in the main text and Sec. I, we can replace operators $\hat{\bf r}$ and $\hat{\bf p}$ in $\partial\hat{H}/\partial{\bf r}$ and $\partial\hat{H}/\partial{\bf p}$ with the complex variables ${\bf r}_c$ and ${\bf p}_c$:
\begin{eqnarray}
    \dot{\bf r}_c(t) &\approx&  \frac{\partial\epsilon({\bf p}_c, {\bf r}_c)}{\partial {\bf p}_c} + \frac{\partial \epsilon({\bf p}_c, {\bf r}_c)}{\partial {\bf r}_c} \times {\bf \Omega}, \nonumber\\
    &\approx& \frac{\partial\epsilon({\bf p}_c, {\bf r}_c)}{\partial {\bf p}_c} - \dot{\bf p}_c \times {\bf \Omega},
\end{eqnarray}
which yields the EOM for the complex variable ${\bf r}_c$ in Eq. 1 in the main text in the presence of $\bf \Omega$.

As an example of the complex semiclassical theory in the presence of multiple bands and the Berry phase, we consider the following non-Hermitian quantum model:
\begin{equation}
    \hat{H}= \alpha \sigma^x \hat{p}_x+\beta \sigma^y \hat{p}_y, \label{eq:ham_berry}
\end{equation}
in the presence of a magnetic field $B_z$. $\alpha, \beta \in \mathbb{C}$ and $\sigma^x, \sigma^y, \sigma^z$ are the Pauli matrices. Quantum mechanically, we have:
\begin{eqnarray}
    \hat{H}^2 &=& \alpha^2 \hat{p}_x^2 + \beta^2 \hat{p}_y^2 - \alpha \beta B_z \sigma^z \nonumber \\
    &=& \omega( 2\hat{b}^{\dagger} \hat{a} + 1 - \sigma^z)/2,
\end{eqnarray}
where $\hat{a}= (\alpha\hat{p}_x + i\beta\hat{p}_y)/\sqrt{\omega}$ and $\hat{b}^\dagger= (\alpha\hat{p}_x - i\beta\hat{p}_y)/\sqrt{\omega}$ are ladder operators similar to the main text. $\omega=2\alpha\beta B_z$. Consequently, the spectrum is $\epsilon_n=\pm \sqrt{\omega n}$, $n\in \mathbb{Z}$ and $n\ge 0$, which serves as our benchmark for the complex semiclassical theory.

The semiclassical equations of motion in the presence of the Berry curvature $\Omega(\bf{p})$ are given by \cite{QianNiu2010RMP}:
\begin{eqnarray}
    \dot{\bf{r}} &=& \partial \epsilon / \partial {\bf{p}} - \dot{\bf{p}} \times \Omega(\bf{p}), \nonumber \\
    \dot{\bf{p}} &=& {\bf E} + \dot{\bf r} \times \bf{B}, \label{eq:eomberry}
\end{eqnarray}
where $\bf{E} = -\partial \epsilon / \partial {\bf{r}}$. As mentioned in the main text, the complex semiclassical theory amounts to generalizing the variables to the complex domains. For the non-Hermitian quantum system in Eq. \ref{eq:ham_berry}, we obtain $\epsilon=\pm \sqrt{\alpha^2 p_x^2+\beta^2 p_y^2}$. With $\bf{E}=0$ and the Berry curvature $\Omega(\bf{p}) \propto \bf{p}$, the corresponding complex EOMs give:
\begin{eqnarray}
    \dot{\bf{p}} &=& (\partial \epsilon / \partial {\bf{p}}) \times B_z \hat{z}, \nonumber \\
    \dot{p_x} &=& B_z \partial \epsilon / \partial {p_y} = \beta^2 p_y B_z / \epsilon,  \\ \nonumber
    \dot{p_y} &=& -B_z \partial \epsilon / \partial {p_x}  = -\alpha^2 p_x B_z / \epsilon,
\end{eqnarray}
whose solution is:
\begin{eqnarray}
    p_x &=& z_1 e^{i\omega't} + z_2 e^{-i\omega't}, \nonumber \\
    p_y &=& (z_1 e^{i\omega't} - z_2 e^{-i\omega't})\cdot i\alpha/\beta, \nonumber\\
    \epsilon^2 &=& 4\alpha^2 z_1 z_2, \label{eq:orbit_berry}
\end{eqnarray}
where $\omega'=\alpha\beta B_z/\epsilon$. $z_1$ and $z_2$ are determined by the initial conditions.

Next, we calculate the Berry phase along the orbit. We obtain the left and right spinor states with $\epsilon>0$ at time $t$ from:
\begin{eqnarray}
    H(t) &=& z_1 e^{i\omega't}(\sigma^x+i\sigma^y) + z_2 e^{-i\omega't}(\sigma^x-i\sigma^y) \nonumber \\
    &=& 2\alpha \begin{pmatrix}
      & z_1 e^{i\omega't} \\
    z_2 e^{-i\omega't} &
\end{pmatrix},
\end{eqnarray}
and evaluate the accumulated Berry phase via integrating over the connection:
\begin{equation}
    \int_0^{\frac{2\pi}{\omega'}} dt \cdot i (\sqrt{z_1 z_2}, z_1 e^{i\omega't}) \partial_t \begin{pmatrix}
    \sqrt{z_1 z_2}\\
    z_2 e^{-i\omega't}
    \end{pmatrix} / 2z_1z_2 = \pi.
\end{equation}
Note that we have employed the biorthogonal convention for expectation values; see Sec. VB. $2z_1z_2$ is a normalization. Consequently, the quantization condition demands:
\begin{eqnarray}
\oint {\bf p}\cdot d{\bf r} &=& \oint -{\bf p} \times d{\bf p} \cdot \hat{z}/B_z = 2\pi z_1 z_2\cdot 2\alpha/\beta B_z,    \nonumber \\
&=& 2\pi(n+1/2)-\pi =2\pi n, \; n\in \mathbb{Z},
\label{eq:berry_qc}
\end{eqnarray}
where we have made use of $d{\bf p} =d{\bf r} \times B_z \hat{z}$ from the EOMs for the first equality. Putting Eq. \ref{eq:berry_qc} and Eq. \ref{eq:orbit_berry} together, we establish the energy spectrum from the complex semiclassical theory:
\begin{equation}
\epsilon=\sqrt{2\alpha\beta B_z n}=\sqrt{\omega n},
\end{equation}
and similarly for the $\epsilon<0$ states. Overall, the results based on the complex semiclassical theory in the presence of the Berry phase are fully consistent with the quantum benchmark.

\end{document}